\documentclass[onecolumn]{aastex631}

\submitjournal{ApJL}

\begin{document}

\title{High optical to X-ray polarization ratio reveals Compton scattering in BL Lacertae's jet}

\author[0000-0002-3777-6182]{Iv\'an Agudo}
\affiliation{Instituto de Astrof\'{i}sica de Andaluc\'{i}a, IAA-CSIC, Glorieta de la Astronom\'{i}a s/n, E-18008 Granada, Spain}

\author[0000-0001-9200-4006]{Ioannis Liodakis}
\affiliation{NASA Marshall Space Flight Center, Huntsville, AL 35812, USA}
\affiliation{Institute of Astrophysics, Foundation for Research and Technology - Hellas, Voutes, 7110, Heraklion, Greece}


\author[0000-0002-4241-5875]{Jorge Otero-Santos}
\affiliation{Instituto de Astrof\'{i}sica de Andaluc\'{i}a, IAA-CSIC, Glorieta de la Astronom\'{i}a s/n, E-18008 Granada, Spain}
\affiliation{Istituto Nazionale di Fisica Nucleare, Sezione di Padova, 35131 Padova, Italy}

\author[0000-0001-9815-9092]{Riccardo Middei}
\affiliation{Space Science Data Center, Agenzia Spaziale Italiana, Via del Politecnico snc, I-00133 Roma, Italy}
\affiliation{INAF Osservatorio Astronomico di Roma, Via Frascati 33, 00078 Monte Porzio Catone (RM), Italy}
\affiliation{Center for Astrophysics \textbar ~Harvard \& Smithsonian, 60 Garden Street, Cambridge, MA 02138 USA}

\author[0000-0001-7396-3332]{Alan Marscher}
\affiliation{Institute for Astrophysical Research, Boston University, 725 Commonwealth Avenue, Boston, MA 02215, USA}

\author[0000-0001-9522-5453]{Svetlana Jorstad}
\affiliation{Institute for Astrophysical Research, Boston University, 725 Commonwealth Avenue, Boston, MA 02215, USA}
\affiliation{Saint Petersburg State University, 7/9 Universitetskaya nab., St. Petersburg, 199034 Russia}

\author[0000-0001-9826-1759]{Haocheng Zhang}
\affiliation{University of Maryland Baltimore County, Baltimore, MD 21250, USA}
\affiliation{NASA Goddard Space Flight Center, Greenbelt, MD 20771, USA}

\author[0000-0003-3556-6568]{Hui Li}
\affiliation{Theoretical Division, Los Alamos National Laboratory, Los Alamos, NM 87545, USA}

\author[0000-0002-5614-5028]{Laura Di Gesu}
\affiliation{ASI - Agenzia Spaziale Italiana, Via del Politecnico snc, 00133 Roma, Italy}

\author[0000-0001-6711-3286]{Roger W. Romani}
\affiliation{Department of Physics and Kavli Institute for Particle Astrophysics and Cosmology, Stanford University, Stanford, California 94305, USA}

% ...



\begin{abstract}
Blazars, supermassive black hole systems (SMBHs) with highly relativistic jets aligned with the line of sight, are the most powerful long-lived emitters of electromagnetic emission in the Universe. We report here on a radio to gamma-ray multiwavelength campaign on the blazar BL Lacertae with unprecedented polarimetric coverage from radio to X-ray wavelengths. The observations caught an extraordinary event on 2023 November 10-18, when the degree of linear polarization of optical synchrotron radiation reached a record value of 47.5\%. In stark contrast, the Imaging X-ray Polarimetry Explorer (IXPE) found that the X-ray (Compton scattering or hadron-induced) emission was polarized at less than 7.4\% (3$\sigma$ confidence level). We argue here that this observational result rules out a hadronic origin of the high energy emission, and strongly favors a leptonic (Compton scattering) origin, thereby breaking the degeneracy between hadronic and leptonic emission models for BL Lacertae and demonstrating the power of multiwavelength polarimetry to address this question. Furthermore, the multiwavelength flux and polarization variability, featuring an extremely prominent rise and decay of the optical polarization degree, is interpreted for the first time by the relaxation of a magnetic ``spring'' embedded in the newly injected plasma. This suggests that the plasma jet can maintain a predominant toroidal magnetic field component parsecs away from the central engine.
\end{abstract}



\section{Introduction}\label{int}

The most remarkable properties of blazars, the most powerful class of active galactic nuclei, whose relativistic jets point at a small angle to the line of sight \citep{blandford2019}, are produced by light-travel effects and relativistic aberration of their broadband radiation. While it is well established that the low-energy component of the spectral energy distribution (SED) of blazars is dominated by non-thermal synchrotron radiation, the nature of their high-energy emission component is a pressing question in high-energy astrophysics. Models based on both leptonic \citep{maraschi1992,dermer1993} and hadronic \citep{aharonian2000} origins of this radiation have been proposed, however without solid evidence to rule out one or the other \citep[see][and references therein]{Hovatta2019}. The latter requires the presence of highly relativistic protons accelerated in the jet. If hadronic processes were confirmed, SMBH jets would be the most plausible candidates for the origin of the recently observed extragalactic high-energy neutrinos \citep{Icecube2018} and ultra high-energy cosmic rays \cite[e.g.,][]{Amaterasu2023}. This would offer new ways for studying fundamental particle physics in extreme regimes. Simultaneous broadband polarimetry from radio wavelengths to X-rays, which is now possible thanks to the novel capabilities of the Imaging X-ray Polarimetry Explorer \citep[IXPE,][]{Weisskopf2022_ixpe_technical}, can finally settle this debate, as different models predict different behaviors of the low- and high-energy polarized emission \cite[e.g.,][]{zhang2013,peirson2019,Peirson2022}.

Blazars are prime candidates for such studies. In particular, those blazars where the emission of the low-energy component peaks at infrared frequencies, namely low- (LSP) and intermediate-synchrotron-peaked (ISP) blazars \citep{abdo2010}, are ideal targets for investigating the origin of the high-energy emission, as their X-ray emission arises mainly from the high-energy component. 
BL Lacertae (BL Lac), the archetype of the BL Lac subclass of blazars, is often classified as LSP, with $\nu_{\text{syn}}\approx4 \times 10^{13}$~Hz \citep[see][]{ajello2020,lott2020}. However, during some flaring states it becomes an ISP blazar ($\nu_{\text{syn}}> 10^{14}$~Hz).
BL Lac has been the target of three previous IXPE observations \citep{middei2023,peirson2023}, two in the LSP and one in the ISP state. Thus far, polarization in the IXPE 2-8 keV band remains undetected, although during the ISP state observed in November 2022, \citet{peirson2023} reported significant polarization of $\sim$22\% in the 2-4 keV sub-band.

\section{Results}\label{res}

\begin{figure}
\centering
\includegraphics[scale=0.5]{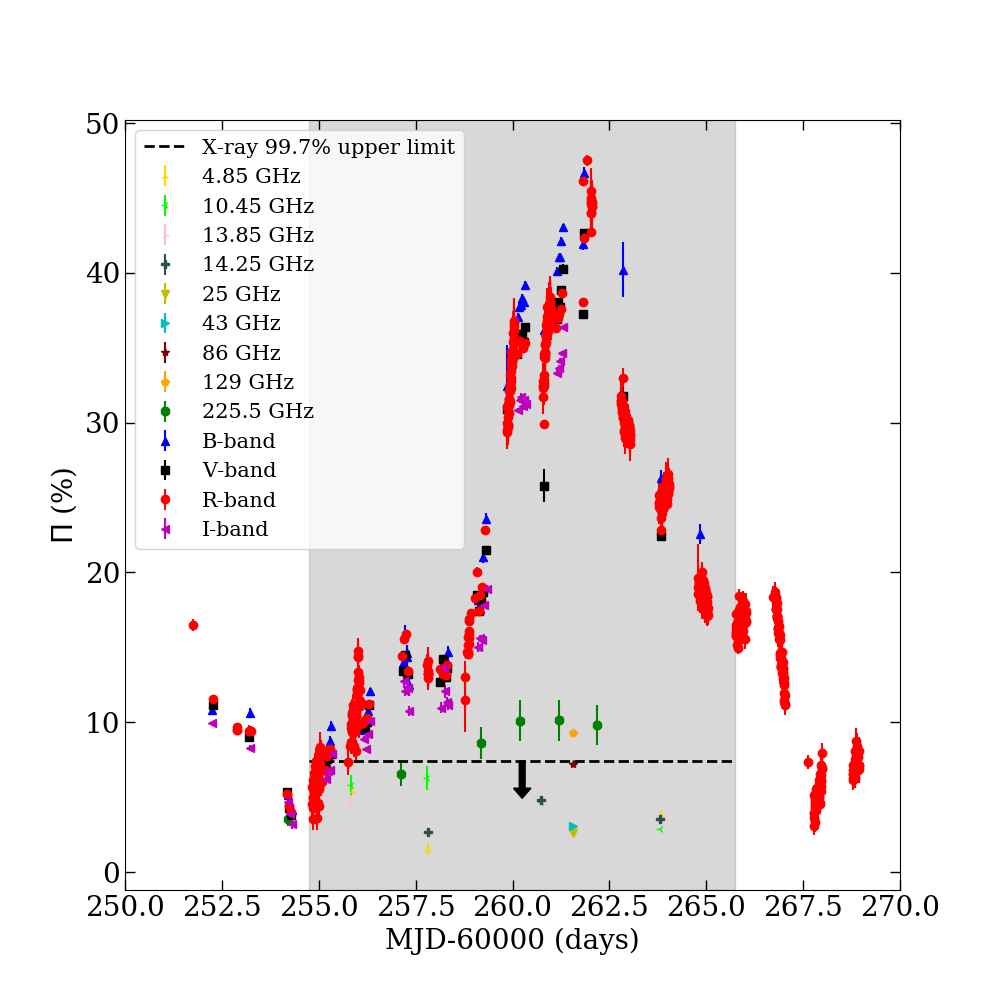}
 \caption{Multiwavelength polarization measurements during the IXPE observing campaign reported here. The grey shaded area marks the duration of the IXPE observation. The different radio and millimeter observing wavelengths and optical bands are marked with different symbols and colors, as shown in the legend. The 99.73\% (3$\sigma$) upper limit of the entire integrated IXPE observation is marked with the black horizontal dashed line and arrow. The error bars correspond to the 68\% (1$\sigma$) confidence interval of every measurement. MJD 60250 corresponds to November 2, 2023.}
\label{plt:pd_all}
\end{figure}

\begin{figure}
\centering
\includegraphics[scale=0.085]{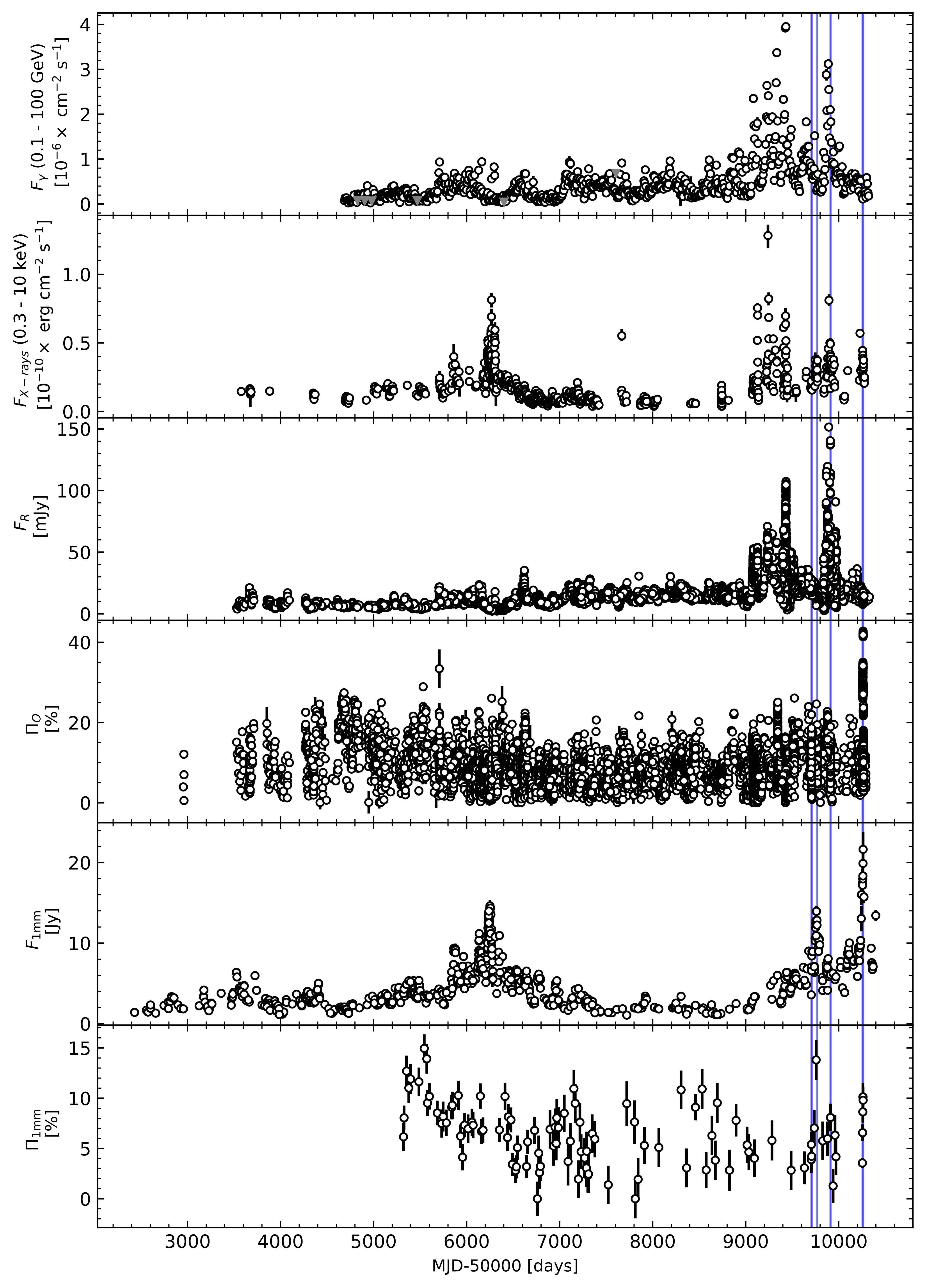}
 \caption{Historical light curves of BL Lacertae between 2005 and 2024. From top to bottom: $\gamma$-ray photon flux between 0.1~GeV and 100~GeV (flux points with black open markers and upper limits with grey markers), X-ray flux between 0.3 and 10~keV, optical $R$-band flux density, optical $R$-band polarization degree (\%), radio 1 mm flux density, and radio 1 mm polarization degree. Blue vertical lines indicate the previous and current IXPE observations of BL Lacertae. A description of the origin of the data in this figure is provided in the Appendix.}

\label{plt:optical_historical}
\end{figure}

BL Lac was observed by IXPE for approximately 500 ks during 2023 November 7-17.
These observations were made during a low state of emission in the $\gamma$-ray range, although the flux in the synchrotron dominated spectral ranges was elevated. In particular, at short mm wavelengths the flux was at the highest level ever reported, while the optical and X-ray fluxes were at moderately high levels; see Figure \ref{plt:optical_historical}.

Several ground and space-borne telescopes in the radio, millimeter, optical, and X-ray bands observed BL Lac during this IXPE observation; see the Appendix, where details about the instruments, observing dates, and data reduction and analysis are provided. Using several telescopes across the world, we were able to achieve nearly uninterrupted optical polarization coverage throughout the IXPE observation. No significant X-ray polarization was detected in the 2-8 keV band, similar to the previous IXPE observations. We are able to place a 99.73\% (3$\sigma$) confidence level upper-limit on the polarization degree of $\rm\Pi_X<7.4\%$ in that band (2-8 keV),  see the Appendix. 
A recent analysis of our IXPE data \citep{2025ApJ...978...43M} also resulted in a non detection of the linear polarization. \citet{2025ApJ...978...43M} gave a less constraining upper limit of $\rm\Pi_X<7.5\%$
at 99\% confidence level.
Our $\rm\Pi_X<7.4\%$ upper limit therefore constitutes the most stringent constraint on the polarization degree from an LSP/ISP blazar, and almost a factor of two improvement over previous IXPE observations of BL Lac. The presence of variability in the polarization direction would lead to incoherent averaging of the polarization vectors, partially canceling any polarization signal \citep[e.g.,][]{DiGesu2023}. We tested for such variability of the X-ray polarization using four independent methodologies; see Section \ref{time-binned-IXPE-polarization}: Time-binned IXPE polarization analysis. None of the tests provided high significance ($>>3\sigma$) evidence of rotation of the polarization angle; see the Appendix.

During the IXPE observation, the measured optical polarization reached a maximum level of $\rm\Pi_O=47.5\pm0.4\%$ (Fig. \ref{plt:pd_all}). This is the most highly polarized state ever reported for BL Lac, and rivals the maximum measured in any blazar \citep[e.g.,][]{Smith2017}. The short millimeter wavelength (1.3 mm, 225.5 GHz) polarization increased to $\rm\Pi_{mm}\sim10\%$, following a similar trend as at optical wavelengths, both in polarization degree and angle. During the optical polarization flare, we also observed total flux brightening in the mm, optical, and X-ray bands (see Fig. \ref{plt:XRT} and the Appendix).
The infrared (IR) to optical and $\gamma$-ray spectral indices ($-1.39\pm0.01$ and $-1.31\pm0.08$, respectively) displayed similar values during the IXPE pointing, while the X-ray spectral index ($-0.68\pm0.01$) lay between the millimeter-IR ($-0.36\pm0.08$) and the IR-optical ($-1.39\pm0.01$) values; see Fig.~\ref{plt:broadband_spectral_indices} and subsection {Broad-band spectral index analysis} in the Appendix. The latter implies that the IR spectral index matches the X-ray slope over part of the IR wavelength range.

\begin{figure}
\centering
\includegraphics[scale=0.08]{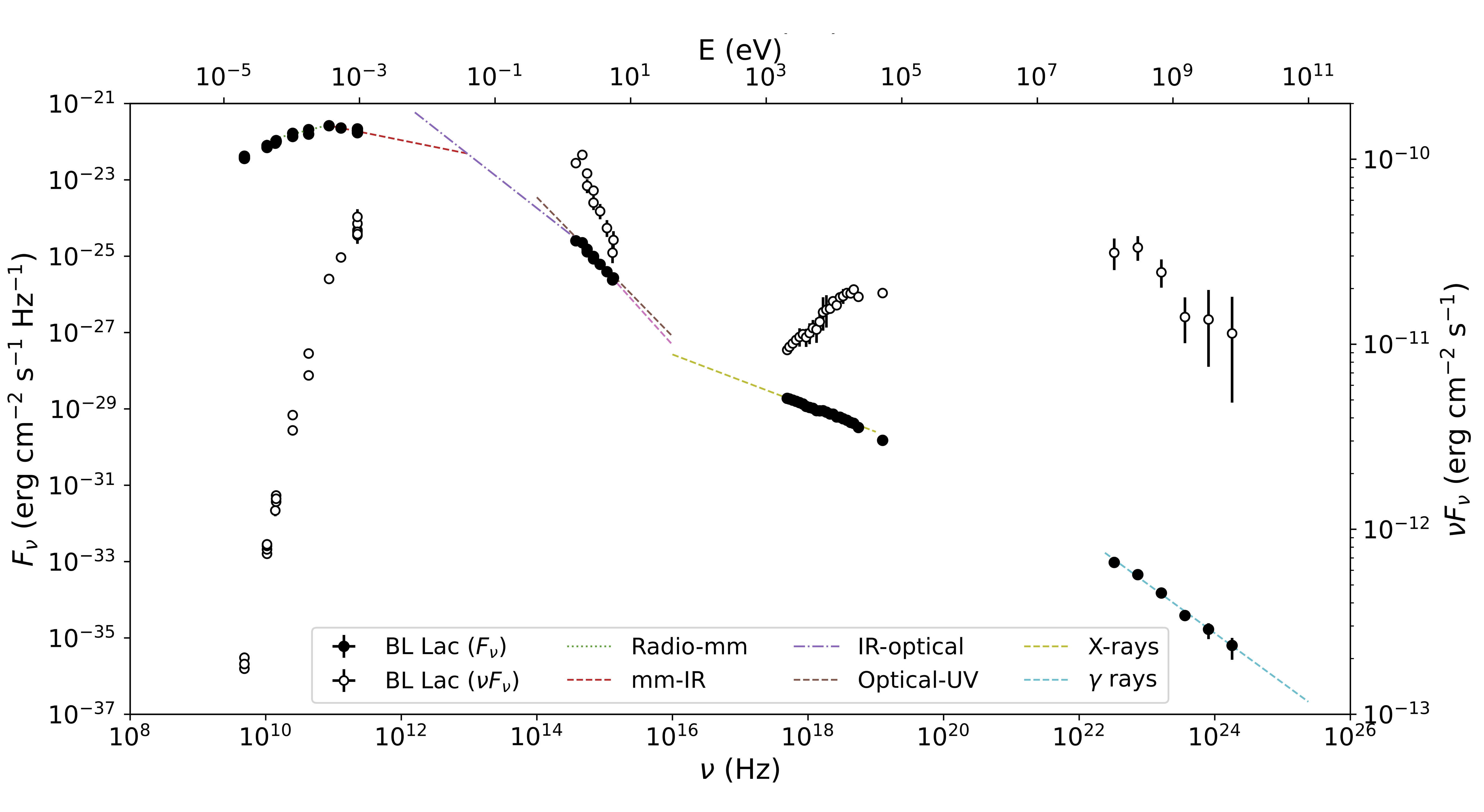}
 \caption{ Flux density $F_{\nu}$ (black markers, in units of erg~cm$^{-2}$~s$^{-1}$~Hz$^{-1}$) at frequency $\nu$ (Hz) and SED (open markers, in units of erg~cm$^{-2}$~s$^{-1}$) of BL Lac from radio to $\gamma$-ray frequencies during the time span of the IXPE observations. The different colored lines correspond to the power-law fits used for the estimation of the spectral indices ($\alpha$, where $F_{\nu}\propto\nu^{\alpha}$) of each band, as specified in the legend and in the Appendix, Table~\ref{tab:spectral_index_analysis}.}

\label{plt:broadband_spectral_indices}
\end{figure}

\section{Discussion}\label{dis}

\begin{table}
\centering
\caption{Summary of the multiwavelength polarization expectations for different X-ray emission scenarios outlined in the text.}
\begin{tabular}{ccc}
\hline
Model & X-ray polarization & X-ray polarization\\
 & degree & variability \\
\hline
single-zone leptonic (EC) & Unpolarized & ---  \\
single-zone leptonic (SSC) & 0.5$\rm\times\Pi_O$ & similar to optical  \\
energy-stratified/multizone leptonic (SSC) & $\rm\leq0.3\times\Pi_O$ & similar to millimeter  \\
single-zone hadronic &  $\rm\approx\Pi_O$ & less variable than optical  \\
energy-stratified/multizone hadronic (flare) &  $\rm\approx\Pi_O$ & less variable than optical  \\
energy-stratified/multizone hadronic (non-flare) & $\rm\approx\Pi_{mm}$ & similar to millimeter  \\
\hline
\end{tabular}
\label{tab:scenarios}
\end{table}

The origin of the high-energy bump of the SED of blazar jets depends on the nature of the radiating particles. 
In the leptonic scenario, electrons (and also positrons, if any) dominate the emission via Compton scattering. For isotropic seed photons, radiation scattered by relativistic leptons in the jet is unpolarized \citep{Bonometto1973}. For an internal photon field (i.e., synchrotron self-Compton, SSC), if the emission is produced in a single localized region in the jet, the polarization of scattered radiation would be $\sim0.5$ times that of the seed photon field \citep{Nagirner1993,Poutanen1994,krawczynski2012}, and the spectral index of the seed photon field should equal that of the scattered photon population. If there is energy stratification in the emission region or there are multiple emission regions, the polarization degree would be lowered by a factor of $\sim3$ or even more  \citep{Liodakis2019,peirson2019,Zhang2024}.
It is also possible that energetic photons are produced 
by scattering by a beam of cold electrons in relativistic bulk motion. This would produce higher polarization in the X-ray than in the optical bands \citep{Begelman1987}. This scenario is rejected by our observations, and also by previous IXPE results \citep{middei2023,peirson2023,Marshall2023,Kouch2024-II}.

In the hadronic scenario, protons can contribute to the high-energy radiation through (i) direct proton-synchrotron radiation or (ii) a photo-pair process that produces electrons and positrons (and other, short-lived, charged leptons), which in turn emit synchrotron radiation. 
If the process is localized in a single emission region, the X-ray polarization would be comparable to that in the optical bands \citep{zhang2013,Paliya2018,Zhang2019}. Even if there is energy stratification in the emission region or there are multiple emission regions, the polarization during a simultaneous multiwavelength flare (i.e., with no pronounced cross-frequency time delays owing to radiative energy losses) is expected to be similar to the case of a single emission region \citep{Zhang2024}. 
 If the contributing protons have a spatially extended distribution inside the jet, the X-ray polarization of the emission from the protons would be comparable to the  millimeter polarization \citep{Zhang2024}. The expected polarization  of the models is summarized in Table \ref{tab:scenarios}. 
 
 Given that protons have much longer cooling timescales than electrons, the X-ray polarization for 
 proton emission processes should show less prominent variability than the optical polarization, which is due to relativistic electrons that lose energy more rapidly through emission of radiation \citep{Paliya2018,Zhang2019}. 
As we do not observe significant variability in the millimeter or optical polarization angle, no significant
 depolarization is expected when averaging over the duration of the IXPE observation.
We find that both the millimeter and optical polarization fractions are higher than the 3$\sigma$ X-ray upper limit during a significantly elevated total flux emission state at radio, optical, and X-ray bands.
The optical polarization is $>6$ times higher than the X-ray upper limit, hence we can reject all of the single-zone hadronic models. 
Based on the expectations of X-ray polarization at least as strong as that at optical or millimeter wavelengths,
the multi-region hadronic scenario can be confidently rejected as well. 

The average (over the IXPE exposure) optical polarization of $25\%$, which is a factor of $\sim3$ lower than that corresponding to a uniform magnetic field, can result from $\sim$9 turbulent cells \citep[e.g.,][]{Marscher2022}. In an energy-stratified shock model, the polarization should depend on frequency as $\nu^{-1/2}$ \citep{Marscher2022}. The expected polarization of the energy-stratified/multi-zone SSC and external Compton (scattering of seed photons originating outside of the jet) emission scenarios is then between 0 and 5\%. This is consistent with the IXPE upper limit. Focusing on the optical polarization peak (MJD~60261-60262.75), for an average optical polarization degree $\rm\Pi_O=40\%$, the SSC expectation is 14\% \cite[see][]{Zhang2024}. Repeating the IXPE analysis within the time interval near the peak yields a 3$\sigma$ upper limit of $<$21\%, again consistent with the SSC scenario.

The optical polarization degree $\Pi_o$ observed during the IXPE observation showed a dramatic increase from an initially low level, followed by a rapid decay, which provides direct evidence for a reconfiguration of global magnetic fields in the jet at $\sim1$ pc from the central engine. Here we suggest a ``magnetic spring'' scenario as the primary mechanism accounting for this drastic variation in polarization. Before the rise in optical polarization, $\Pi_o$ was $\sim5\%$, with the polarization angle aligned along the jet direction. As jets typically possess both poloidal and toroidal magnetic components, this suggests that the toroidal magnetic field was comparable to, or weaker than, poloidal and/or turbulent magnetic field components. At its peak ($\Pi_o = 47.5 \pm 0.4\%$), the polarization angle remained aligned with the jet direction, indicating that the toroidal component became at least 2.5 times stronger than the poloidal and/or turbulent components. We suggest that this could be caused by an injection of a magnetic ``spring'' with a dominant toroidal configuration, embedded in the jet plasma flowing outward from the central engine. Such an injection could arise from a magnetic eruption from a magnetically arrested disk \citep[e.g.,][]{Ripperda2022} and it continues into the relativistic jet \citep{Yang2024}. Subsequently, the rise and fall of the optical polarization over $\sim2$ weeks imply a rapid relaxation of the injected toroidal component in a compact region ($\sim$ 0.1 pc) such that the toroidal component eventually becomes comparable to the poloidal components, as could occur through a kink instability in the jet \citep{Guan2014,Barniol2017}. The relaxation time of the toroidal component via kink instabilities is on the order of the Alf\'ven crossing time, which is about a week for typical blazars \citep{Zhang2017}. The observed approximately symmetric rise and fall in $\Pi_o$ suggest that magnetic fields in this jet play an active role in restoring their configurations, different from a low-magnetization plasma environment where the rise and fall in $\Pi_o$ are not expected to be symmetric \citep{Zhang2016}. With a kink instability, some of the toroidal magnetic field can be converted into the poloidal component, reducing $\Pi_o$ and contributing to the curved jet morphology observed in VLBA images (Figure \ref{plt:vlbi_maps}).

The injection of a toroidally-dominated magnetic spring can also account for the multiwavelength flaring behavior. This injection leads to a sudden increase of the total magnetic field strength, producing the historically high flare in the mm-band. This flare occurs near the low-energy spectral turnover, with synchrotron radiation emitted by relatively low-energy electrons (compared to those emitting at optical and X-ray wavelengths), whose cooling times are on the order of years. The fact that the mm-band polarization degree does not rise as much as the optical band is consistent with the scenario that mm-band and lower energy emission comes from a much larger volume, subject to contributions from many local magnetic field configurations and probably turbulent magnetic fields in the jet, as well as synchrotron self-absorption effects.  
This is consistent with the energy stratified picture reported in previous X-ray polarization observations of HSPs \citep{marscher1985,liodakis2022,DiGesu2023,middei2023,Kouch2024}. This is also supported by a BL Lac analysis made by \citet{2025ApJ...978...43M} on a similar time span using a less populated dataset without optical and radio polarization information. 
Meanwhile, the optical band, near the high-energy turnover of the synchrotron SED, shows suppressed flaring due to enhanced synchrotron cooling (and therefore less efficient acceleration) of high-energy electrons. This is consistent with the observed softening of the optical spectral index (Table \ref{tab:spectral_index_analysis}). Similarly, there is no prominent increase in $\gamma$-ray flux, since it is produced mainly by Compton scattering by the same electrons that emit optical synchrotron radiation.

\section{Conclusion}\label{con}
Our results unambiguously reject single-zone hadronic processes and multizone/energy stratified hadronic scenarios in BL Lac. Instead, they favor a scenario where X-ray emission is dominated by Compton-scattered emission from relativistic electrons in a region with gradients in the maximum electron energy. This could be realized where electrons are accelerated in a fresh plasma blob threaded by a magnetic spring dominated by a strong toroidal magnetic field in the parsec-scale region, beyond which electrons are subsequently advected downstream. While protons may occasionally play an important role in the emission from blazar jets, and their interactions in jets might be associated with neutrino detections, our polarization study implies that this is not the typical case. Our observations constitute the strongest test of X-ray emission processes in astrophysical jets so far, and the strongest evidence against relativistic protons dominating the X-ray emission in blazars.
Our study also demonstrates the rich potential of multiwavelength total flux and polarimetric monitoring campaigns that include X-ray polarimetry and ultra-high angular resolution VLBI observations to resolve the long-standing question of the role that relativistic protons play in blazar jets.
  
\section*{Acknowledgments}
The Imaging X-ray Polarimetry Explorer (IXPE) is a joint US and Italian mission.  The US contribution is supported by the National Aeronautics and Space Administration (NASA) and led and managed by its Marshall Space Flight Center (MSFC), with industry partner Ball Aerospace (contract NNM15AA18C)---now, BAE Systems.  The Italian contribution is supported by the Italian Space Agency (Agenzia Spaziale Italiana, ASI) through contract ASI-OHBI-2022-13-I.0, agreements ASI-INAF-2022-19-HH.0 and ASI-INFN-2017.13-H0, and its Space Science Data Center (SSDC) with agreements ASI-INAF-2022-14-HH.0 and ASI-INFN 2021-43-HH.0, and by the Istituto Nazionale di Astrofisica (INAF) and the Istituto Nazionale di Fisica Nucleare (INFN) in Italy. This research used data products provided by the IXPE Team (MSFC, SSDC, INAF, and INFN) and distributed with additional software tools by the High-Energy Astrophysics Science Archive Research Center (HEASARC), at NASA Goddard Space Flight Center (GSFC). 
Based on observations obtained with XMM-Newton, an ESA science mission with instruments and contributions directly funded by ESA Member States and NASA. 
We acknowledge the use of public data from the Swift data archive.
Some of the data are based on observations collected at the Observatorio de Sierra Nevada; which is owned and operated by the Instituto de Astrof\'isica de Andaluc\'ia (IAA-CSIC); and at the Centro Astron\'{o}mico Hispano en Andaluc\'ia (CAHA); which is operated jointly by Junta de Andaluc\'{i}a and Consejo Superior de Investigaciones Cient\'{i}ficas (IAA-CSIC). 
The Perkins Telescope Observatory, located in Flagstaff, AZ, USA, is owned and operated by Boston University.
Data from the Steward Observatory spectropolarimetric monitoring project were used. This program is supported by Fermi Guest Investigator grants NNX08AW56G, NNX09AU10G, NNX12AO93G, and NNX15AU81G. This research was partially supported by the Bulgarian National Science Fund of the Ministry of Education and Science under grants KP-06-H38/4 (2019) and KP-06-PN-68/1(2022). The Liverpool Telescope is operated on the island of La Palma by Liverpool John Moores University in the Spanish Observatorio del Roque de los Muchachos of the Instituto de Astrofisica de Canarias with financial support from the UKRI Science and Technology Facilities Council (STFC) (ST/T00147X/1). This research has made use of data from the RoboPol program, a collaboration between Caltech, the University of Crete, IA-FORTH, IUCAA, the MPIfR, and the Nicolaus Copernicus University, which was conducted at Skinakas Observatory in Crete, Greece.
The data in this study include observations made with the Nordic Optical Telescope, owned in collaboration by the University of Turku and Aarhus University, and operated jointly by Aarhus University, the University of Turku and the University of Oslo, representing Denmark, Finland and Norway, the University of Iceland and Stockholm University at the Observatorio del Roque de los Muchachos, La Palma, Spain, of the Instituto de Astrofisica de Canarias. The data presented here were obtained in part with ALFOSC, which is provided by the Instituto de Astrof\'{\i}sica de Andaluc\'{\i}a (IAA) under a joint agreement with the University of Copenhagen and NOT.
The Submillimeter Array (SMA) is a joint project between the Smithsonian Astrophysical Observatory and the Academia Sinica Institute of Astronomy and Astrophysics and is funded by the Smithsonian Institution and the Academia Sinica. Maunakea, the location of the SMA, is a culturally important site for the indigenous Hawaiian people; we are privileged to study the cosmos from its summit. 
The POLAMI observations reported here were carried out at the IRAM 30m Telescope. IRAM is supported by INSU/CNRS (France), MPG (Germany) and IGN (Spain). 
The KVN is a facility operated by the Korea Astronomy and Space Science Institute. The KVN operations are supported by KREONET (Korea Research Environment Open NETwork) which is managed and operated by KISTI (Korea Institute of Science and Technology Information). 
Partly based on observations with the 100-m telescope of the MPIfR (Max-Planck-Institut f\"ur Radioastronomie) at Effelsberg. Observations with the 100-m radio telescope at Effelsberg have received funding from the European Union's Horizon 2020 research and innovation programme under grant agreement No 101004719 (ORP). 
The IAA-CSIC co-authors acknowledge financial support from the Spanish "Ministerio de Ciencia e Innovaci\'{o}n" (MCIN/AEI/ 10.13039/501100011033) through the Center of Excellence Severo Ochoa award for the Instituto de Astrof\'{i}isica de Andaluc\'{i}a-CSIC (CEX2021-001131-S), and through grants PID2019-107847RB-C44 and PID2022-139117NB-C44. 
I.L was supported by the NASA Postdoctoral Program at the Marshall Space Flight Center, administered by Oak Ridge Associated Universities under contract with NASA.
B. A.-G., I.L were funded by the European Union ERC-2022-STG - BOOTES - 101076343. Views and opinions expressed are however those of the author(s) only and do not necessarily reflect those of the European Union or the European Research Council Executive Agency. Neither the European Union nor the granting authority can be held responsible for them.
J.O.-S. acknowledges financial support from the project ref. AST22\_00001\_9 with founding from the European Union - NextGenerationEU, the \textit{Ministerio de Ciencia, Innovaci\'on y Universidades, Plan de Recuperaci\'on, Transformaci\'on y Resiliencia}, the \textit{Consejer\'ia de Universidad, Investigaci\'on e Innovaci\'on} from the \textit{Junta de Andaluc\'ia} and the \textit{Consejo Superior de Investigaciones Cient\'ificas}, as well as from INFN Cap. U.1.01.01.01.009.
This work has been partially supported by the ASI-INAF program I/004/11/4. 
The research at Boston University was supported in part by National Science Foundation grant AST-2108622, NASA Fermi Guest Investigator grants 80NSSC23K1507 and 80NSSC23K1508, NASA NuSTAR Guest Investigator grant 80NSSC24K0547, and NASA Swift Guest Investigator grant 80NSSC23K1145. This work was supported by NSF grant AST-2109127. 
HZ is supported by NASA under award number 80GSFC21M0002. HZ's work is supported by Fermi GI program cycle 16 under the award number 22-FERMI22-0015, and IXPE GO program cycle 1 under the award numbers 80NSSC24K1160 and 80NSSC24K1173.
Research by HL was supported by the Laboratory Directed Research and Development program of Los Alamos National Laboratory under project number  20220087DR. E. L. was supported by Academy of Finland projects 317636 and 320045. We acknowledge funding to support our NOT observations from the Finnish Centre for Astronomy with ESO (FINCA), University of Turku, Finland (Academy of Finland grant nr 306531).
S. Kang, S.-S. Lee, W. Y. Cheong, S.-H. Kim, and H.-W. Jeong  were supported by the National Research Foundation of Korea (NRF) grant funded by the Korea government (MIST) (2020R1A2C2009003). 
CC acknowledges support by the European Research Council (ERC) under the HORIZON ERC Grants 2021 programme under grant agreement No. 101040021.
This work was supported by JST, the establishment of university fellowships towards the creation of science technology innovation, Grant Number JPMJFS2129.  
D.B. acknowledge support from the European Research Council (ERC) under the European Unions Horizon 2020 research and innovation program under grant agreement No.~771282.

%

\vspace{5mm}
\facilities{IXPE, \textit{Swift}(XRT and UVOT), XMM-Newton, NuSTAR, \textit{Fermi},
Belogradchik Observatory, Calar Alto Observatory, Nordic Optical Telescope, Liverpool Telescope, LX-200, Perkins Telescope, Sierra Nevada Observatory, Skinakas observatory, SMA, IRAM 30m Telescope, KVN, Effelsberg 100-m Telescope, VLBA.}


\software{HEASoft \citep{2014ascl.soft08004N}, FTOOLS \citep{1995ASPC...77..367B}, IOP4 \citep{escudero2023,escudero2024}, AIPS\footnote{\url{http://www.aips.nrao.edu/aipsdoc.html}}, DIFMAP \citep{1997ASPC..125...77S}, GILDAS\footnote{\url{https://www.iram.fr/IRAMFR/GILDAS/}}}




\appendix\label{sec11}
The IXPE observation of BL Lac presented here was accompanied by contemporaneous observations from several ground and space-based facilities. We give a short description for each of those facilities below. More details of the observations and data reduction, including information on the individual telescopes, can be found in \cite{liodakis2022,digesu2022,peirson2023,middei2023,Kouch2024}.

\section{\textit{Fermi}-LAT $\gamma$-ray data}
Data from the pair-conversion Large Area Telescope (LAT) on board the \textit{Fermi} satellite are included in this work. This telescope monitors the $\gamma$-ray sky every three hours in the energy range of 20~MeV to $\sim$2~TeV, with sensitivity highest in the 0.1--200 GeV range \citep{atwood2009}. 
We have retrieved the publicly available BL Lac light curve from the \textit{Fermi}-LAT public Light Curve Repository \citep[LCR\footnote{\url{https://fermi.gsfc.nasa.gov/ssc/data/access/lat/LightCurveRepository/about.html}}, see][]{abdollahi2023}. This database contains the light curves of all variable sources in the 4FGL-DR2 catalogue \citep{abdollahi2020,ballet2020}, which includes sources with a variability index $>$21.6 \citep[see Table 12 in][]{abdollahi2020}. 
We refer to \cite{abdollahi2023} for a detailed discussion on the automatic unbinned likelihood analysis adopted by the LCR for computing the light curves. Here, we have used the 3-day binned light curve with a freely varying spectral slope in each bin. Figure \ref{plt:optical_historical} shows the  $\gamma$-ray lightcurve compiled since the beginning of operations of \textit{Fermi}. The light curve shows that the $\gamma$-ray flux level of the source during the IXPE observation reported here was low to moderate.

We have derived the $\gamma$-ray portion of the SED and spectrum of BL~Lac during the IXPE observation by retrieving and analyzing the data from the LAT Data Server\footnote{\url{https://fermi.gsfc.nasa.gov/ssc/data/access/lat/}}, plus an extra $\pm$15 days around it, i.e., between 2023 October 23 and 2023 December 3. This extended time window was selected owing to the relatively faint and stable $\gamma$-ray emission of the source in the period that is strictly simultaneous with IXPE observations, which would lead only to upper limits over that period. We have selected and analyzed all of the Pass8 \texttt{P8R3\_source} events between 100~MeV and 300~GeV with version 1.2.23 of the standard software \textsc{Fermitools}. A cut of 90$^{\circ}$ in zenith angle was applied to minimize the contamination from the limb of the Earth. The recommended Galactic diffuse emission model and isotropic component for the event selection and analysis performed here were employed\footnote{\url{https://fermi.gsfc.nasa.gov/ssc/data/access/lat/BackgroundModels.html}}.

In order to obtain a model describing all of the relevant sources in the field, a binned likelihood analysis was performed on 1 year of LAT data (January 2019 to January 2020). We included all sources contained in the defined region-of-interest (ROI) near BL Lac, plus those located within an additional annular region of 10$^{\circ}$ radius. For inclusion into the final model, the spectral parameters and flux normalization of all sources within the ROI were left as free parameters. For the sources within the annular region, these parameters were fixed to those from the 4FGL-DR2 catalogue. The normalization of the diffuse components were also left as free parameters. In the convergence iteration, all sources with a test statistic (TS) $<$4 (approximately 2$\sigma$ significance) were removed from the final model.

Finally, we have also extracted the $\gamma$-ray spectrum and SED of BL Lac in the aforementioned time window. We used a binned likelihood analysis to model each bin of the SED with a power-law spectral model. The LAT catalogues show a significant preference for a log-parabolic spectral shape after averaging more than 14 years of data. However, due to the short interval considered here, we do not accumulate enough statistics to observe a preference for a curved spectrum over the simpler power law model. Hence, we adopted the latter for the SED estimated here, obtaining a power-law spectrum with an index $\alpha=-2.31 \pm 0.08$.

\section{X-ray data}

\subsection{IXPE observation and data reduction} 

BL Lac was observed by IXPE's three Detector Units (DUs) from 2023-11-07 00:44:08 to 2023-11-17 19:48:05 for a total exposure time of $\sim$515 ks. The $I$, $Q$, and $U$ spectra for each of the three DUs of IXPE were computed using the software \textsc{ixpeobssim} \citep[v. 31.0.1,][]{pesce2019,baldini2022} and adopting the background rejection prescriptions by \citet{DiMarco2023}. The spectra were computed to enable the use of the weighted analysis method \citep[][]{dimarco2022}. The Stokes $I$, $Q$, and $U$ spectra were extracted using a circular region with radius=0.95$'$ centered on BL Lac. The background was derived using an annular region with $r_{\rm in(out)}$=1.2$'$(3.5$'$). These choices have been shown to enhance the sensitivity to polarization, as discussed by \citet{DiMarco2023}. The resulting $I$ Stokes spectra were rebinned, requiring each energy bin to have a signal-to-noise ratio greater than 7. A uniform binning of 280 eV was then adopted for the $Q$ and $U$ Stokes spectra.\\

\subsection{Swift, XMM-Newton and NuSTAR observations and data reduction}

\begin{table}
	\centering
	\caption{Log of X-ray observations related to the IXPE pointings of BL Lac.}\label{log}
	\begin{tabular}{c c c c}
		\hline
		Observatory & Obs. ID & Obs. date & Net exp. \\
		& & yyyy-mm-dd & ks \\
		\hline 
		{IXPE} & 02009701  & 2023/11/07-17 &$\sim$ 515\\
		{NuSTAR} & 80901639002  & 2023/11/13-14 & $\sim$ 20\\
		{XMM-Newton} & 0902112601& 2023/11/17&$\sim$5\\
         \hline 
	\end{tabular}
\end{table}

\begin{figure}
\centering
\includegraphics[width=\textwidth]{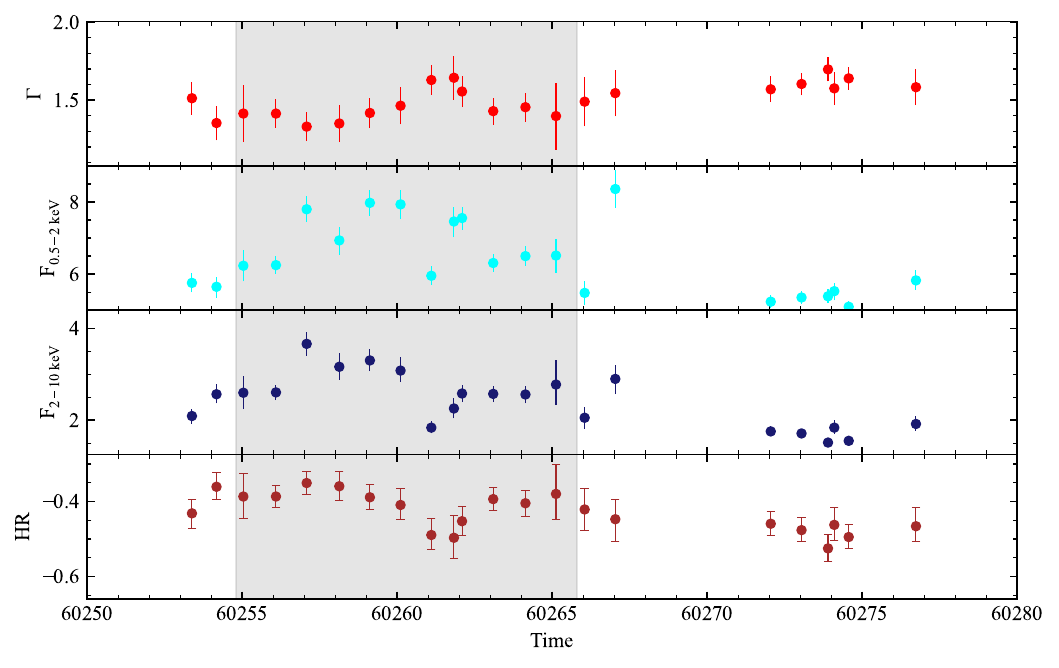}
\caption{Results of the Swift-XRT monitoring campaign of BL Lac. Fluxes F$_{\rm 0.5-2~keV}$, and F$_{\rm 2-10~keV}$ are displayed in units of $10^{-12}$ erg cm$^{-2}$ s$^{-1}$ and $10^{-11}$ erg cm$^{-2}$ s$^{-1}$, respectively. The label ``HR'' stands for hardness ratio, which we define as (F$_{\rm 2-10~keV}$-F$_{\rm 0.5-2~keV}$)/(F$_{\rm 2-10~keV}$+F$_{\rm 0.5-2~keV}$). Gray shaded areas mark the duration of the IXPE pointing.}
 \label{plt:XRT}
\end{figure}

\indent The IXPE observation of BL Lac was coordinated  with different X-ray observatories: XMM-Newton \citep[][]{Jansen2001}, the Neil Gehrels Swift observatory \citep[Swift,][]{Gehrels2004}, and the Nuclear Spectroscopic Telescope Array \citep[NuSTAR,][]{Harrison2013}. This joint effort allowed us to monitor the source variability and derive the intrinsic spectral shape of the source below and above the IXPE 2-8 keV bandpass.\\
\indent The X-ray Telescope (XRT) on {\it Swift} operates in the 0.2-10 keV energy band and has a high spatial resolution of 18$''$. Both the XRT and UltraViolet and Optical Telescope (UVOT) operate simultaneously to enable concurrent observations in different electromagnetic bands \citep{Rom05,Bur05}. We retrieved the  reducedX-ray, UV, and optical data from the {\it Swift} public mirror archive \footnote{Available at \url{https://swift.ssdc.asi.it/}} of the Space Science Data Center (SSDC) at the Italian Space Agency (ASI). 
{\it Swift}-XRT observed BL Lac in Photon Counting (PC) readout mode. The data were first reprocessed locally with the {\footnotesize XRTDAS} software package (version {\ttfamily v3.7.0}), developed by the ASI-SSDC and included in the NASA-HEASARC {\footnotesize HEASoft} package\footnote{Available at \url{https://heasarc.gsfc.nasa.gov/docs/software/heasoft/}} (version {\ttfamily v6.31.1}). Standard calibration and filtering processing steps were adopted and the calibration files available from the {\itshape Swift}-XRT {\footnotesize CALDB} (version {\ttfamily 20220803}) were used. The science products were extracted using a circle of 20 pixels (47$''$) radius centered on the target source, while the background was derived from a nearby circular region with a radius of 40 pixels. All the spectra were subsequently binned with the {\ttfamily grppha} tool of the {\footnotesize FTOOLS} package\footnote{Available at \url{https://heasarc.gsfc.nasa.gov/ftools/}} requiring each bin to have a minimum of 5 counts, and then modeled using the {\footnotesize XSPEC} software package\footnote{Available at \url{https://heasarc.gsfc.nasa.gov/xanadu/xspec/}} while adopting a single power-law model with foreground photoelectric absorption from gas in our Galaxy.
In Figure~\ref{plt:XRT} we show the temporal behavior of the power-law slope (photon index), soft and hard fluxes (0.5-2 and 2-10 keV, respectively) and hardness ratios of BL Lac. \\
\indent The UVOT on board Swift is capable of observations in the $[170,600]$\,nm band performed simultaneously with Swift-XRT. In our case, the UVOT pointed at BL Lac with filters V 5468\AA, B 4392\AA, U 3465\AA, UVW1 2600\AA, UVM2 2246\AA, UVW2 1928\AA. We derived the photometric points for each filter for the different observations using the standard tools within an automated reduction procedure. We adopted two regions, a circular one centred on the source (radius =
2$''$) for estimating the counts from the target, while a concentric annulus ($\Delta$radius = 7$''$), free
of any other sources or spurious detection, was used to determine the background. We show the returned count rates for the UVOT filters, uncorrected for Galactic or intrinsic reddening, in Figure~\ref{plt:XRT}.

\indent XMM-Newton, which is sensitive to soft X-rays, briefly observed BL Lac for about 5 ks, from 2023-11-17 12:21:37 to 2023-11-17 19:08:46. The satellite's EPIC-pn camera \citep[][]{Struder2001} observed BL Lac in Small Window mode, with the medium filter applied. Data were processed with the standard XMM-Newton Science Analysis System \citep[SAS v21][]{Gabriel2004}. 
Source extraction radii and screening for high-background intervals were determined through an iterative process \citep{Piconcelli2004} that maximizes the signal-to-noise ratio. The background was extracted from circular regions with a radius of 50$''$, and the same shape centered on BL Lac was adopted for science products. The resulting third-level products were grouped by requiring each bin to contain at least 30 counts, and not to over-sample the spectral resolution by a factor larger than 3. The net count rate was less than the maximum allowed limit of 50 cts s$^{-1}$ to avoid deteriorated response due to photon pile-up for EPIC-pn observations in Small Window mode \citep[e.g.][]{Jethwa2015}. We further assessed the potential impact of pile-up in the XMM-Newton observation by means of the \emph{epatplot} task, a standard SAS command devoted to checking for any pile-up affecting the data, and we found it to be negligible.

\indent NuSTAR observed BL Lac simultaneously with IXPE for a total of $\sim$20 ks from 2023-11-13 16:11:09 to 2023-11-14 02:51:09. The FPMA/B instruments carried on its focal plane operate in the 3-79 keV range, thereby providing unique broadband data. Science products of BL Lac were obtained by employing the {\footnotesize NuSTARDAS} software, also developed by the ASI-SSDC and included in the NASA-HEASARC {\footnotesize HEASoft} package, and the latest calibration data files {\footnotesize CALDB} (version {\ttfamily 20220510}). We used a circular region (radius r$\sim$70'') centered on BL Lac to extract both the FMPA and FPMB spectra. The same sized region, but located on a black area of the detectors, was used to extract the background. The ancillary and response matrices were computed at this stage, and we grouped the spectra via the grppha standard command in order to have at least 30 counts per bin.

\subsection{Time-averaged spectro-polarimetric analysis}

\begin{figure}
\centering
\includegraphics[width=\textwidth]{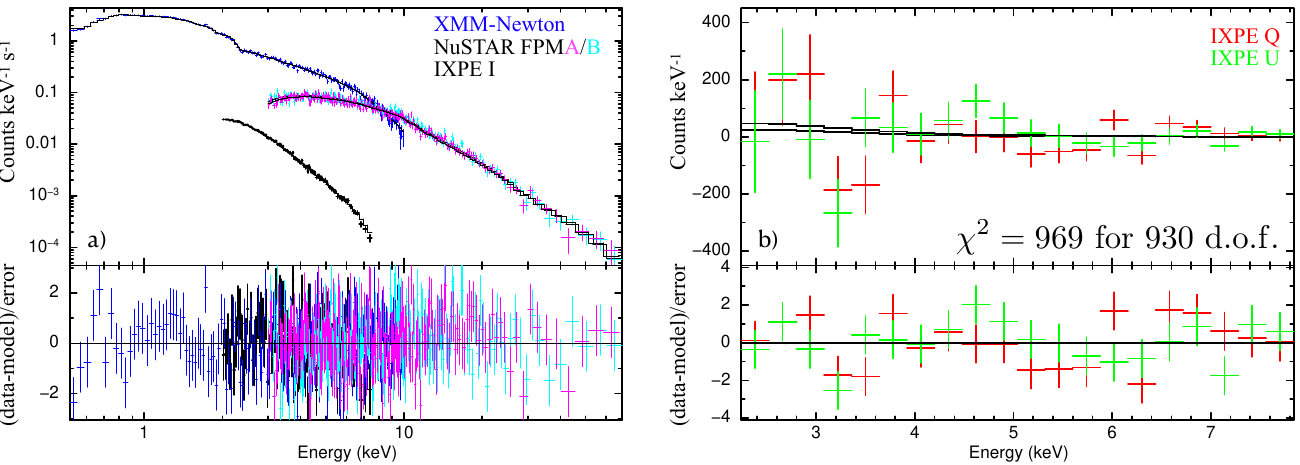}
\includegraphics[scale=0.5]{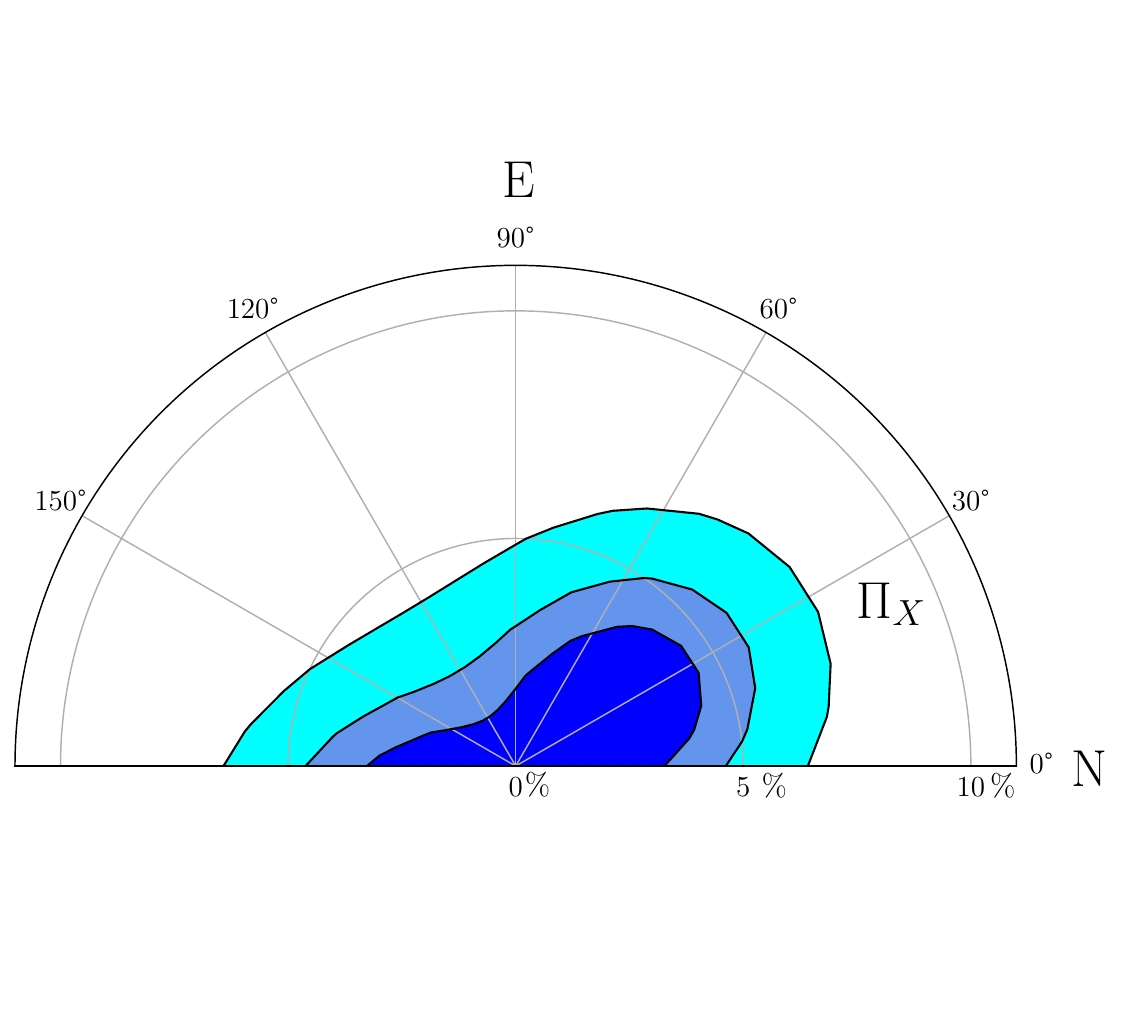} \caption{Top panels: Best fit to the IXPE, XMM-Newton, and NuSTAR data and corresponding residuals. On the left, the model \textit{tbabs$\times$const$\times$polconst$\times$po} fitting the $I$ Stokes spectra is shown, while the right panel displays the best fit $Q$ and $U$ Stokes spectra. Bottom panel: confidence regions corresponding to 68\%, 90\%, and 99\% uncertainties derived using the full IXPE 2-8 keV data.}
\label{bestfit}
\end{figure}

\indent 
Our search for a polarized X-ray signal from BL Lac relied on a spectro-polarimetric analysis in which we simultaneously modeled the IXPE, XMM-Newton, and NuSTAR spectra. We proceeded in the manner of \citet{middei2023}, where a similar dataset was analysed. In particular, we tested a simple phenomenological model in {\it XSPEC} that can be written as \texttt{tbabs$\times$const$\times$po$\times$polconst}.
The first component accounts for the column density through gas in our Galaxy and whatever additional hydrogen might intercept our line of sight \citep[e.g.][]{Bania1991,Madejski1999,middei2023}. The constant is used for the inter-calibration of the different telescopes, and the power law describes the X-ray continuum spectrum. The polarization signal is encoded in the $Q$ and $U$ Stokes spectra, and is determined from the multiplicative component {\it polconst}, which assumes the X-ray polarized signal to be constant in the IXPE operating band.\\
\indent The spectro-polarimetric fit is performed by computing the photon index of the continuum and its normalisation,
the polarization degree $\Pi_{\rm X}$ and polarization angle $\Psi_{\rm X}$, and five constants accounting for the DU1, DU2 and DU3 detector units of IXPE, the EPIC-pn camera, and the FPMA/B modules of NuSTAR. These simple steps returned a fit with $\chi^2$/d.o.f.=1039/932. Similarly to the discussion in \citet{middei2023}, the observed column density exceeds the Galactic value and the EPIC-pn spectrum shows bump-like residuals around 0.7 keV that may be due to some additional spectral component. 
We investigated the possible origin of these bump-like residuals by examining whether the unmodeled data could be attributed to a steep upper tail of the synchrotron spectrum entering the XMM-Newton band pass. To test this, we replaced the Gaussian line by a second power law for which the photon index and the normalization were fitted. However, this attempt returned a poorer fit with $\chi^2$/d.o.f.=638/553. As a second test, we replaced the two power laws with a broken power law (\textsc{bknpo} in XSPEC notation) and fit the two $\Gamma$ the break energy and the normalization of the power law. This second attempt yielded to a better fit ($\chi^2=$556 for 553 d.o.f.) but the very flat soft spectral shape predicted by this model ($\Gamma_{\rm soft}\sim0.2$) is not consistent with the hypothesis of a soft synchrotron component extending up to $\sim$2 keV.
We thus proceeded as in \citet{middei2023} and updated our fitting model to include an unpolarized {\it apec} component that could explain the excess in the soft X-ray band as emission from hot diffuse plasma. The addition of this Gaussian component is beneficial in terms of $\chi^2$, as the statistic improved by $\Delta\chi^2$/$\Delta$d.o.f.$=-70/-2$. This best-fitting model is shown in Figure~\ref{bestfit}, and the inferred quantities are reported in Table ~\ref{ixpefit}.
Our analysis reveals that the X-ray spectrum of BL Lac beyond $\sim 2$ keV was well represented by a power law with photon index $\Gamma=1.87\pm$0.01. The density of the absorbing column, although exceeding the Galactic value, was found to be $N_{\rm H}$=2.45$\pm$0.05$\times10^{21}$ cm$^{-2}$, in agreement with previous studies \citep[e.g.][]{Weaver2020}. Below the IXPE bandpass, the EPIC-pn spectrum seems compatible with emission from hot gas ($kT=0.38\pm$0.04 keV), although a physical origin for such a hot gas is unknown and requires further study. \\
\indent According to our analysis, no significant polarization of BL Lac was measured by IXPE during the campaign. The 3$\sigma$ (99.73\% confidence level) upper limit to the polarization degree is $\Pi^{3\sigma}_{\rm X}<$7.4\%, while the corresponding value at 99\% confidence level is $\Pi^{99}_{\rm X}<$6.6\%. No indications of significant polarization could be found by altering the energy range. Finally, all cross-normalization constants are in agreement with each other within $\sim$20\%.

\begin{table}
\centering
	\caption{Best-fit parameters for the joint X-ray observations. All uncertainties are quoted at 68\% confidence level for one parameter of interest. The 2-8 keV flux is estimated from the IXPE data, while fluxes in the 0.5-2 and 2-10 keV bands are derived from the XMM-Newton data.}\label{ixpefit}
	\begin{tabular}{c c c c}
	\hline
	component & parameter& value & units\\
	\hline
    polconst& $\rm \Pi_X$ &$<7.4\%$ (C.L. 99.73\%) &-\\
     & $\rm \Psi_X$ & -&-\\
    tbabs& N$_{\rm H}$ &2.45$\pm$0.07 &$\times$ 10$^{21}$ cm$^{-2}$\\
    powerlaw& $\Gamma$ & 1.87$\pm0.01$&-\\
    & Norm &$6.72\pm0.01$ & $\times$10$^{-3}$ photons keV$^{-1}$ cm$^{-2}$ s$^{-1}$\\
    const &k$_{\rm DU2}$&$0.97\pm0.01$ &-\\
     const &k$_{\rm DU3}$&$0.88\pm0.01$ &-\\
     const &k$_{\rm pn}$&$0.78\pm0.01$ &-\\
     const &k$_{\rm FPMA}$&$1.07\pm0.02$ &-\\
     const &k$_{\rm FPMB}$&$1.11\pm0.02$ &-\\
    \hline
    \hline
    $\rm F_{\rm 2-8~keV}$&& $1.75\pm0.01$&$\times$10$^{-11}$ erg cm$^{-2}$ s$^{-1}$\\
    $\rm F_{\rm 2-10~keV}$&& $1.60\pm0.02$&$\times$10$^{-11}$ erg cm$^{-2}$ s$^{-1}$\\
    $\rm F_{\rm 0.5-2~keV}$&& $6.45\pm0.05$&$\times$10$^{-12}$ erg cm$^{-2}$ s$^{-1}$\\

    \hline
		\end{tabular}

\end{table}

\subsection{Time-binned IXPE polarization analysis}
\label{time-binned-IXPE-polarization}

\begin{figure}[t!]
\centering
\includegraphics[width=0.43\hsize]{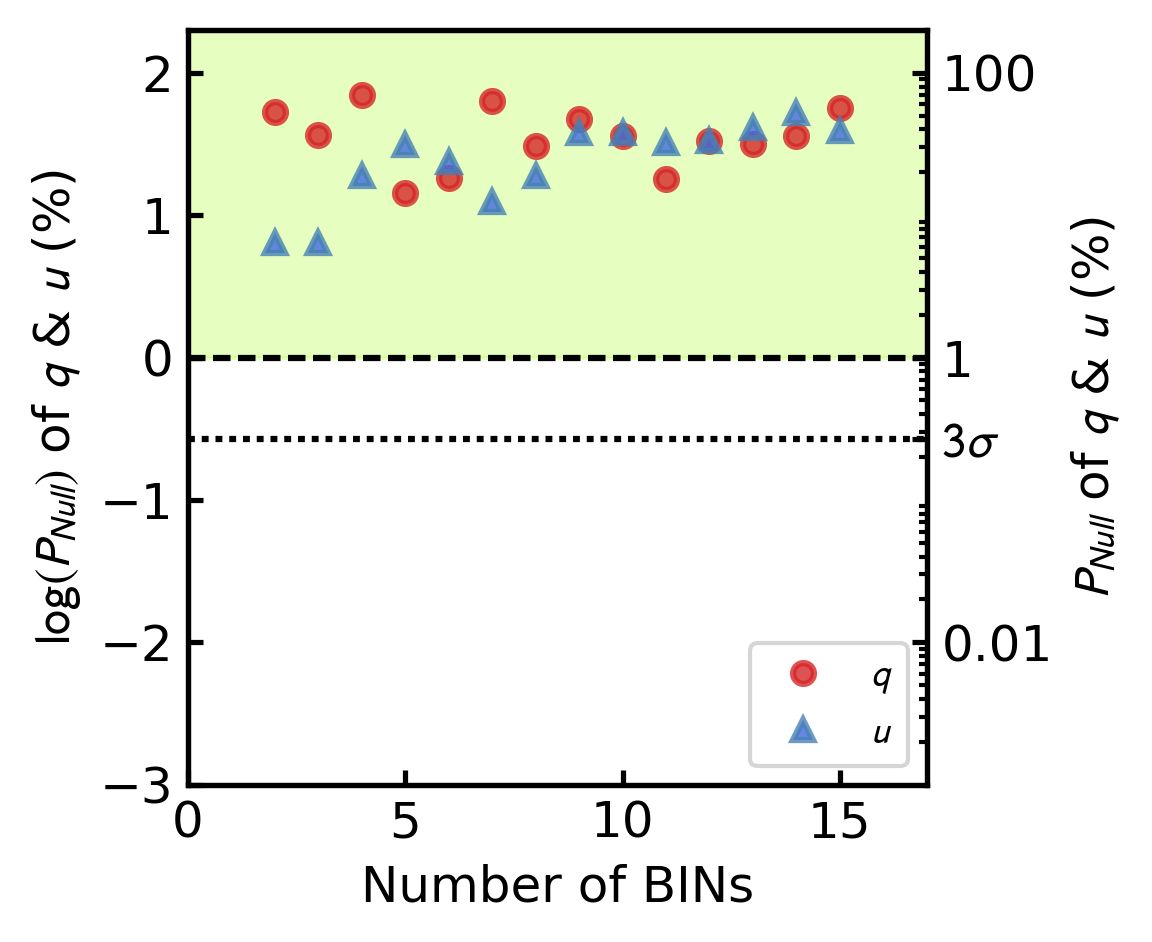}
\quad
\includegraphics[width=0.43\hsize]{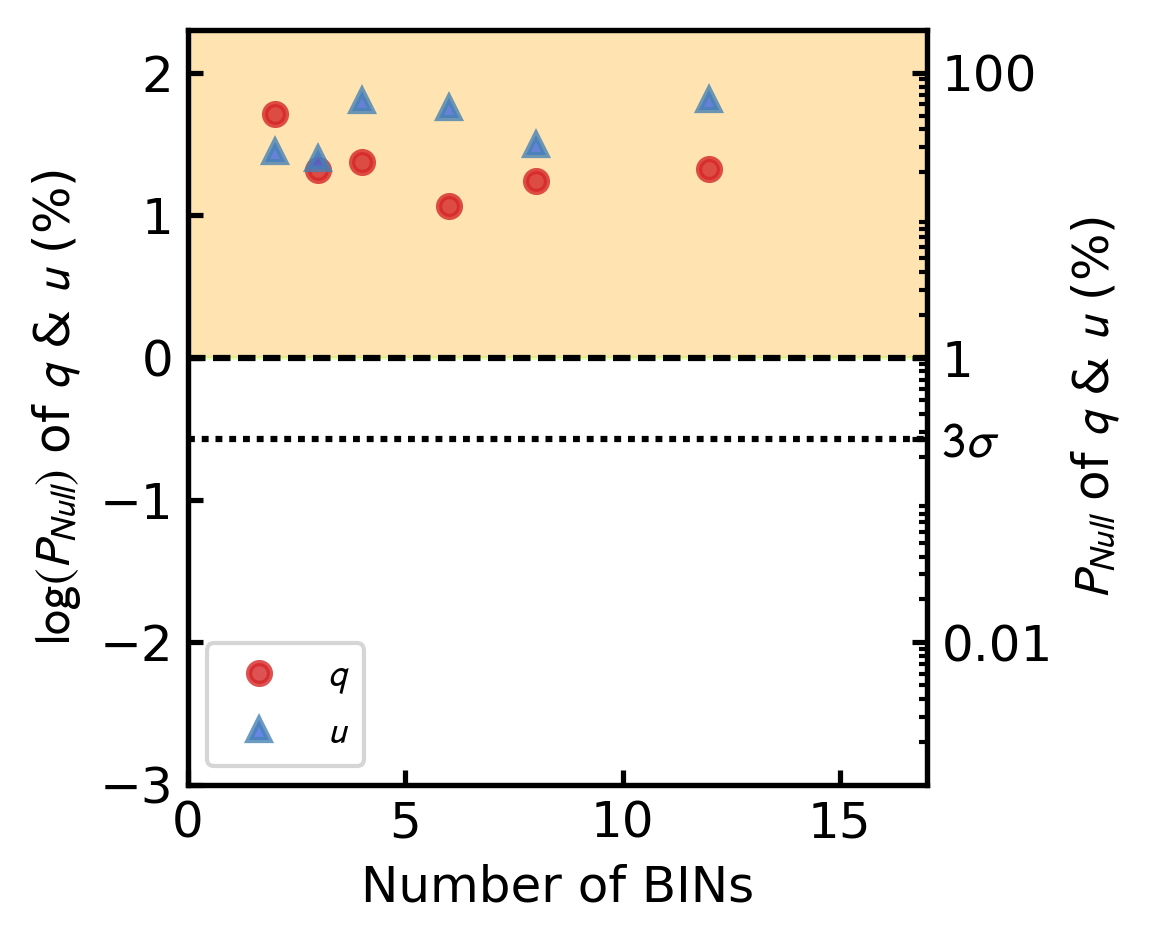}
\caption{
Null hypothesis probability of a constant model of the $q$ (red) and $u$ (blue) Stokes parameters for different time (left panel) and energy (right panel) bins. The green and orange shaded areas indicate that the null hypothesis probability is above the 1\% threshold level. The black dashed and dotted lines located in the middle of each panel represent 1\% and 3$\sigma$ (0.27\%) probability, respectively. The left and right vertical axes of each panel correspond to the probability values in logarithmic and linear scales in percentage, respectively.
}
\label{fig:variability}
\end{figure}

We examined polarization variability over time and energy using a $\chi^2$ test by dividing the entire observation into specific time and energy bins, as described in \cite{2024A&A...681A..12K}. The $\chi^2$ analysis was performed to estimate the null hypothesis probability ($P_{Null}$) of a constant model fit to the time-averaged $q$ and $u$ values of each bin obtained from the independent spectral modeling analysis. In the case of time variability, we estimated the time-averaged $q$ and $u$ values by dividing the entire observation period into specific sub-periods after dividing the data into two to 15 bins (e.g., 2 bins = 515ks/2 bins = 258ks per bin). To test for energy dependence, we divided the entire energy band (2-8 keV) into smaller energy bins, such as two energy bins (2-4 and 4-8 keV), three (2-4, 4-6, and 6-8 keV), and so on (up to 12). Figure \ref{fig:variability} illustrates the calculated $P_{Null}$ of $q$ and $u$ in percentage units for all of the different time binning cases. We found no statistically significant change in polarization when dividing the data into narrower time or energy ranges, with a probability higher than the 1\% threshold in all cases.

We furthermore performed a study of the change in polarization over time. The data set was divided into equal intervals, and the polarimetric analysis was performed on each of them. Several interval durations were tested and a compatible result was obtained in each case. As a trade-off, we show the division in 6 time bins that provide higher time resolution and reasonable statistics in each time bin. We treated the background as in the time-averaged analysis and, as before, considered a weighted method. We used both a spectro-polarimetric analysis and the model-independent analysis described in \cite{Kislat2015}.
In the former case, we tested a simple phenomenological XSPEC model that can be written as \texttt{tbabs$\times$constant$\times$pow$\times$polconst}. In the latter, we applied the same method as for the time-averaged analysis. 
Several statistical tests were applied; the unpolarized model could not be rejected above the $\sim 3~\sigma$ level in any bin or under any condition. Given the current sensitivity, no significant polarization detection can be therefore claimed.  We have also performed several tests to look for energy dependence of the polarization. For this purpose, we tested the spectral model with $pollin$ in XSPEC for each time bin. No significant energy dependence of the polarization was found. 

Finally, because rotation of the EVPA during the IXPE observations could affect the measurement of the polarization degree, we have also checked the data for such a rotation. Using a 1-D test for EVPA rotation \citep{DiGesu2023} we found evidence at 99.5\% confidence for a rotation of 12$\pm$3.5$^{\circ}$/day in the 2-8 keV band.  Furthermore, dividing the data into four independent intervals and allowing for EVPA rotation in each interval, $i$, we find that the likelihood that the polarization is zero is rejected at 99.2\% confidence.  This method involves computing 

\begin{equation}
\mathcal{S = \sum_i^N [ S_i(\hat{q_i}, \hat{u_i}, \hat{R_i}) - S_i(0,0,0) ]}
\end{equation}

\noindent
where $N=4$, $\hat{q_i}$, $\hat{u_i}$, and $\hat{R_i}$ are the best fit parameters and $S_i$ is the event-based log-likelihood computed for the interval $i$ \citep{2024ApJ...964...88M,digesu2022}.  Thus, $\mathcal{S}$ is the 3-D likelihood ratio test of the hypothesis that the polarization is zero for all intervals and is distributed as $\chi^2$ with $3N$ degrees of freedom.  Finally, assuming $R=12$ deg/day, the source polarization is 5.57 $\pm$ 1.75 \%.

\section{Optical data}

\begin{figure}
\centering
\includegraphics[scale=0.5]{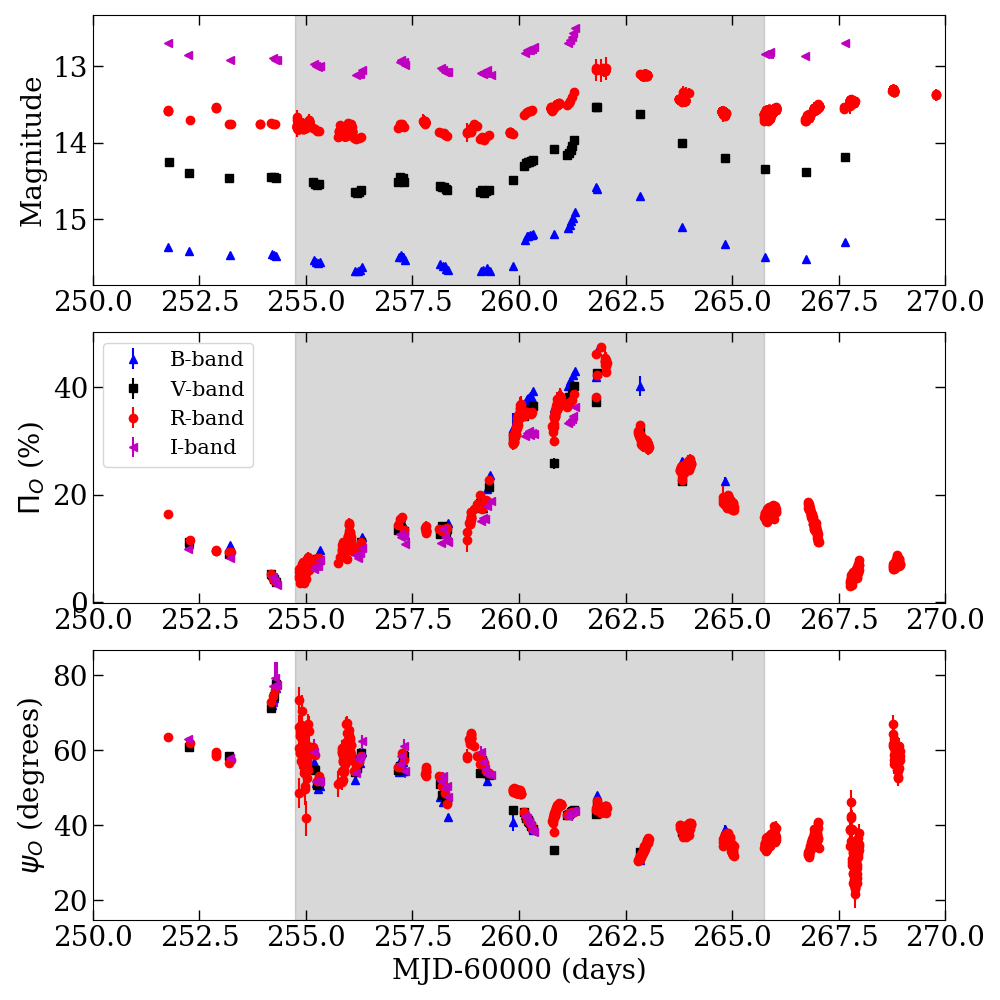}
 \caption{Optical polarization measurements during the IXPE observation. The top panel shows the brightness in magnitudes, the middle panel the polarization degree (\%), and the bottom panel the polarization angle (degrees). The grey shaded area marks the duration of the IXPE observation. The different optical bands are marked with different symbols and colors as shown in the legend. The error bars correspond to the 68\%  (1$\sigma$) confidence interval.}

\label{plt:optical_pol}
\end{figure}

Optical observations covering the entirety of the IXPE exposure were obtained at the Belogradchik Observatory \cite[$R$-band,][]{Bachev2023}, Calar Alto Observatory \cite[CAFOS, $R$-band,][]{escudero2024}, Nordic Optical Telescope \cite[ALFOSC, $R$-band,][]{Nilsson2018}, Liverpool Telescope \cite[MOPTOP, $BVR$-bands,][]{Shrestha2020}, LX-200 photometry ($BVRI$-bands) and polarimetry ($R$-band), Perkins Telescope \cite[PRISM, $BVRI$-bands, photometry and polarimetry,][]{Jorstad2010}, Sierra Nevada Observatory \citep[DIPOL-1, $R$-band,][]{otero2024}, and the Skinakas observatory \cite[RoboPol, $R$-band,][]{Ramaprakash2019}. The CAFOS and DIPOL-1 observations were analyzed using IOP4 \citep{escudero2023,escudero2024}, while the RoboPol observations used the automatic pipeline described in \cite{Panopoulou2015,Blinov2021}. The 40cm LX-200 telescope uses an imaging photo-polarimeter with a ST-7 camera and two Savart plates oriented 45$^\circ$ with respect to each other. The $R$-band polarization observations were corrected for interstellar and instrumental polarization using standard stars.

In the case of BL Lac, as well as many other BL Lac type objects, the host galaxy contributes significantly to the total optical emission. This leads to dilution of the intrinsic polarization degree. To account for this effect, we use the host-galaxy light-distribution model from \cite{Nilsson2007} to estimate the  contribution of the host galaxy to the flux density (in mJy) at the aperture used for the polarization analysis for each observatory. We then correct the observed polarization degree by subtracting the host flux density from the total flux density following \cite{Hovatta2016}. Since we only have a model for the host galaxy in the R-band, only the R-band measurements have been corrected. In this case the correction is small, yielding a change of only 1--3\%. Although we have not corrected the other optical bands, the brightness of a typical elliptical galaxy, as is the case for the host galaxy of BL Lac, falls sharply towards higher frequencies. The host correction in the B- and V-bands is typically negligible, while is it more significant in the I-band. 

The optical polarization measurements are shown in Figure \ref{plt:optical_pol}. We see achromatic variations in brightness and polarization properties. There is a high-amplitude outburst in polarization degree that coincides with the IXPE observation. The polarization degree starts at 4\%, rises to 47.5\%, then declines to 16\% from the beginning to the end of the IXPE observation. This matches the highest polarization ever reported from a blazar (the previous record of $47.4\pm0.1$\% was observed in PKS~1502+106, \citealp{Smith2017}). At the same time, there is a less prominent flare in optical total flux. The polarization angle slowly drifts from 70$^\circ$ to 30$^\circ$ with a median of 45$^\circ$, without any indications of large variations or rotations, as are often exhibited by BL Lac \cite[e.g.,][]{Marscher2008,Blinov2018}. This particular optical behavior of BL Lac will be studied in detail in a forthcoming publication.

In order to provide a general overview of the behavior of BL Lac as compared to the one during this campaign, we have retrieved the historical photo-polarimetric $R$-band data from different observatories and databases. This includes measurements from Steward Observatory\footnote{\url{http://james.as.arizona.edu/~psmith/Fermi}} \citep[2008 to 2018,][]{smith2009}, Skinakas Observatory \citep[RoboPol\footnote{\url{https://robopol.physics.uoc.gr}}, 2013 to 2017,][]{Blinov2021}, Calar Alto Observatory \citep[CAFOS,][]{agudo2012} and Sierra Nevada Observatories (2007 to 2024), Perkins Telescope (2005 to 2023), LX-200 and AZT-8 telescopes (2005 to 2023), Nordic Optical Telescope (NOT, 2015 to 2020), and Tuorla blazar monitoring (KVA, 2003 to 2011), covering a period of more than 20 years. These data are shown in Figure~\ref{plt:optical_historical}. BL Lac is known for its remarkable variability across all bands, as exemplified by the past several bright flares, which reached optical magnitudes as bright as $R \sim11$. Extreme flares have also been observed in X-rays and $\gamma$ rays from this source. However, as shown by the variations in polarization degree over $\sim$20 years of monitoring, the nature of the event observed during the IXPE campaign is unique, reaching an all-time maximum polarization degree for any blazar. Interestingly, this feature coincides with a very low $\gamma$-ray emission state of BL Lac, as shown by the \textit{Fermi}-LAT data, and with a relatively modest optical flare.

\section{Radio/millimetre data}

\begin{figure}
\centering
\includegraphics[scale=0.5]{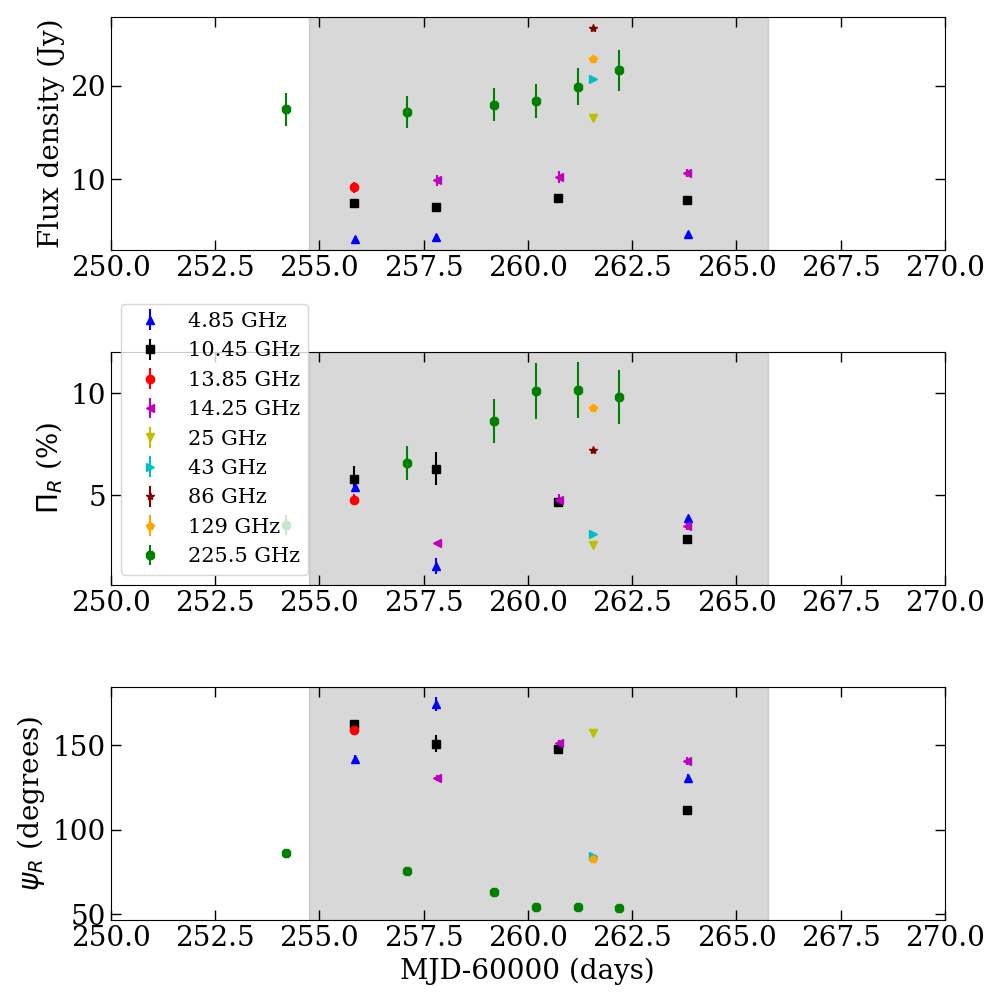}
 \caption{Short-millimetre-wavelength and radio polarization measurements during the IXPE observation. The top panel shows the flux density, the middle panel the polarization degree, and the bottom panel the polarization angle. The grey shaded area marks the duration of the IXPE observation. The different radio frequencies are marked with different symbols and colors as shown in the legend. The error bars correspond to the 68\%  (1$\sigma$) confidence level.}
\label{plt:radio_pol}
\end{figure}

BL Lac was observed from 4~GHz to 225.5~GHz , by the Effelsberg 100-m telescope as part of the Monitoring the Stokes $Q$, $U$, $I$ and $V$ Emission of AGN jets in Radio (QUIVER) program \cite[4.8, 10.4, 13.8, 14.2~GHz,][]{Krauss2003,Myserlis2018}, the Korean VLBI Network  \cite[KVN, 25, 43, 86, 129~GHz,][]{Kang2015}, and by the Submillimeter Array (SMA) within the framework of the SMA Monitoring of AGNs with Polarization (SMAPOL) program \cite[225.5~GHz,][]{Ho2004,Marrone2008,Primiani2016}.

Figure \ref{plt:radio_pol} shows the contemporaneous radio observations. We find a similar behavior as in the optical observations. There is a brightening across frequencies that coincides with a sharp peak in the polarization degree, best visible in the better sampled 225.5~GHz observations. As usual in blazars \cite[e.g.,][]{Agudo2018b}, there is an increase of the polarization degree towards higher frequencies. At low radio frequencies ($<86$~GHz) the polarization degree is in the range 1--6\%. At higher frequencies ($\geq86$~GHz) the polarization degree is between 6--10\%. There is a similar trend in the polarization angle at 225.5~GHz, from 85$^\circ$ to 51$^\circ$, as we observe in the optical bands, although the polarization angle in the radio bands is higher than the optical. The jet axis on the plane of the sky is found to be $10^\circ\pm2^\circ$ \citep{Weaver2022} hence the radio and optical polarization angles wander by tens of degrees from the jet direction. (However, see the discussion of contemporaneous VLBA images below for a different viewpoint.)

We have also extended the 1mm radio light curve with the archival data from SMA available since 2002 \citep{2007ASPC..375..234G} and the Polarimetric Monitoring of AGN at Millimeter Wavelengths \citep[POLAMI, see][]{agudo2018} program running at the 30m IRAM Millimeter Radio Telescope since 2010 (and at 3mm since 2006). The historical 1mm flux density and polarization degree light versus time are displayed in the bottom panels of Figure~\ref{plt:optical_historical}. From the 1mm flux density evolution, it is clear that this IXPE observing window coincides with the brightest total flux ever observed from BL~Lac, with the two previous highest states, corresponding to two fast flares in 2012 and 2022, being a factor of 1.5 weaker than the late 2023 level. Comparison with the archival POLAMI observations reveals that the polarization degree in the 1mm band during this period was above the average value observed for this source.

\begin{figure}
\centering
\includegraphics[scale=0.35]{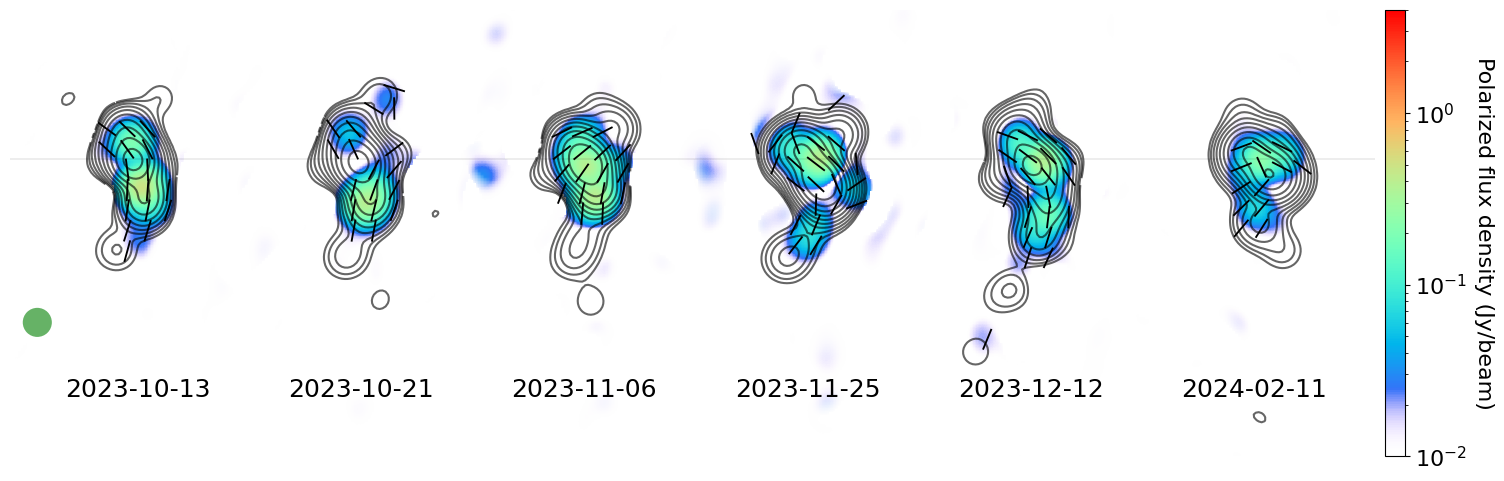}
 \caption{Sequence of 43GHz VLBA images of BL Lac from the BEAM-ME program at 0.1 mas angular resolution. Contours represent total flux density from 0.01 Jy/beam (the lowest contour on every image) to 5.50 Jy/beam (the highest contour). There are 10 logarithmically equally spaced total flux contours from those minimum and maximum levels. The short sticks symbolize the direction of the electric vector polarization angle.}

\label{plt:vlbi_maps}
\end{figure}

We also analyzed a set of six ultra-high-angular resolution VLBA images from the 7mm BEAM-ME program\footnote{See {\url{https://www.bu.edu/blazars/BEAM-ME.html/bllac.html}}} from 2023-10-13 to 2024-02-11; see Figure \ref{plt:vlbi_maps}. The data were reduced, calibrated, imaged and model-fit as in \citet{Jorstad2017}. The 7mm VLBA images show that there is a significant change of polarization angle in the VLBI core between November 6 and 25 from -24$^\circ$ to +53$^\circ$.
In agreement with the sharp total flux emission peak seen in the 1mm light curve (e.g., Fig. \ref{plt:optical_historical}), these two VLBA observing epochs have the highest map peak of 15.14 and 16.24 Jy per beam, respectively, over all BEAM-ME observations started in June 2007. Nnote that the size of the beam is the same.) On 12 December, the polarization angle in the innermost jet region is $\sim40^\circ$, close to the optical median value. Also, the direction of this innermost jet region in Nov. 2023 is 42$^\circ$, very close to the polarization angle in the optical and in the brightest VLBI core emission at 43 GHz on November 24. This behavior is consistent with that expected from a plane-perpendicular shock wave propagating from the innermost bent jet-regions in BL Lac just after the optical total flux and polarization flare reported in Figure \ref{plt:optical_pol}. The close time coincidence of this millimeter-wave and optical behavior strongly suggests that the site of the extreme optical behavior reported in this paper is in the VLBI core region in BL Lac.

\section{Broad-band spectral index analysis}\label{sec:spectral_indices}

We have conducted an evaluation of the spectral indices of the different bands along the electromagnetic spectrum, finding further evidence for the leptonic SSC nature of the high-energy emission observed from BL Lac. For this, we have used the \textsc{python} package \texttt{JetSeT}\footnote{\url{https://jetset.readthedocs.io/en/latest/}} \citep{tramacere2009,tramacere2011,tramacere2020}, which contains an already implemented routine that allows calculation of the spectral ($\alpha$) and photon indices ($\Gamma$, with $\Gamma=\alpha-1$) in different frequency ranges of the electromagnetic spectrum through a power law fit. In particular, we have calculated the spectral indices in the radio to millimeter ($10^{10}-10^{11}$~Hz), millimeter to IR ($10^{11}-10^{13}$~Hz), IR to optical ($10^{12.2}-10^{14.8}$~Hz), optical to UV ($10^{14}-10^{16}$~Hz), X-rays ($10^{16}-10^{19}$~Hz) and MeV to GeV ($10^{22.38}-10^{25.38}$~Hz) bands. The derived spectral and photon index values are reported in Table~\ref{tab:spectral_index_analysis}. 
In Figure~\ref{plt:broadband_spectral_indices} we plot the flux density $F_{\nu}$ (in units of erg~cm$^{-2}$~s$^{-1}$~Hz$^{-1}$) versus frequency, based on the spectral fits performed in each band.

\begin{table}
\centering
\caption{Results of the spectral analysis performed over the different frequency bands, comparing the photon and spectral indices. The last column shows the number of energy bins $N$ used for each fit.}
\label{tab:spectral_index_analysis}
\begin{tabular}{cccccc}
\hline
Band & $\log{\nu_{min} [Hz]}$ & $\log{\nu_{max} [Hz]}$ & Photon index & Spectral index & $N$ \\ \hline
Radio-mm  & 10 & 11 & $-0.55 \pm 0.01$ & $0.45 \pm 0.01$ & 13 \\ 
mm-IR  & 11 & 13 & $-1.36 \pm 0.08$ & $-0.36 \pm 0.08$ & 7 \\ 
IR-optical  & 12.2 & 14.8 & $-2.39 \pm 0.01$ & $-1.39 \pm 0.01$ & 4 \\ 
Optical-UV  & 14 & 16 & $-2.82 \pm 0.03$ & $-1.82 \pm 0.03$ & 10 \\ 
X-rays  & 16 & 19 & $-1.68 \pm 0.01$ & $-0.68 \pm 0.01$ & 23 \\ 
$\gamma$ rays  & 22.38 & 25.68 & $-2.31 \pm 0.08$ & $-1.31 \pm 0.08$ & 6  \\ \hline
\end{tabular}
\end{table}

Our broad-band spectral-index analysis clearly indicates that the optical and $\gamma$-ray spectrum have essentially the same slope, while the X-ray spectral index lies between the millimeter-to-IR and the IR-to-optical values. This is exactly the broad-band spectral behavior expected for an SSC scenario, where the $\gamma$ rays are produced by Compton scattering of the optical synchrotron photons by the same electrons that emit the optical radiation, while the IR synchrotron photon field is the seed for the X-ray SSC emission, upscattered by the IR-emitting electrons.


\bibliography{bibliography}{}

\begin{thebibliography}{}
\expandafter\ifx\csname natexlab\endcsname\relax\def\natexlab#1{#1}\fi
\providecommand{\url}[1]{\href{#1}{#1}}
\providecommand{\dodoi}[1]{doi:~\href{http://doi.org/#1}{\nolinkurl{#1}}}
\providecommand{\doeprint}[1]{\href{http://ascl.net/#1}{\nolinkurl{http://ascl.net/#1}}}
\providecommand{\doarXiv}[1]{\href{https://arxiv.org/abs/#1}{\nolinkurl{https://arxiv.org/abs/#1}}}

\bibitem[{{Abdo} {et~al.}(2010){Abdo}, {Ackermann}, {Agudo}, {Ajello}, {Aller}, {Aller}, {Angelakis}, {Arkharov}, {Axelsson}, {Bach}, {Baldini}, {Ballet}, {Barbiellini}, {Bastieri}, {Baughman}, {Bechtol}, {Bellazzini}, {Benitez}, {Berdyugin}, {Berenji}, {Bland ford}, {Bloom}, {Boettcher}, {Bonamente}, {Borgland}, {Bregeon}, {Brez}, {Brigida}, {Bruel}, {Burnett}, {Burrows}, {Buson}, {Caliandro}, {Calzoletti}, {Cameron}, {Capalbi}, {Caraveo}, {Carosati}, {Casand jian}, {Cavazzuti}, {Cecchi}, {{\c{C}}elik}, {Charles}, {Chaty}, {Chekhtman}, {Chen}, {Chiang}, {Chincarini}, {Ciprini}, {Claus}, {Cohen-Tanugi}, {Colafrancesco}, {Cominsky}, {Conrad}, {Costamante}, {Cutini}, {D'ammando}, {Deitrick}, {D'Elia}, {Dermer}, {de Angelis}, {de Palma}, {Digel}, {Donnarumma}, {Silva}, {Drell}, {Dubois}, {Dultzin}, {Dumora}, {Falcone}, {Farnier}, {Favuzzi}, {Fegan}, {Focke}, {Forn{\'e}}, {Fortin}, {Frailis}, {Fuhrmann}, {Fukazawa}, {Funk}, {Fusco}, {G{\'o}mez}, {Gargano}, {Gasparrini}, {Gehrels}, {Germani}, {Giebels},
  {Giglietto}, {Giommi}, {Giordano}, {Giuliani}, {Glanzman}, {Godfrey}, {Grenier}, {Gronwall}, {Grove}, {Guillemot}, {Guiriec}, {Gurwell}, {Hadasch}, {Hanabata}, {Harding}, {Hayashida}, {Hays}, {Healey}, {Heidt}, {Hiriart}, {Horan}, {Hoversten}, {Hughes}, {Itoh}, {Jackson}, {J{\'o}hannesson}, {Johnson}, {Johnson}, {Jorstad}, {Kadler}, {Kamae}, {Katagiri}, {Kataoka}, {Kawai}, {Kennea}, {Kerr}, {Kimeridze}, {Kn{\"o}dlseder}, {Kocian}, {Kopatskaya}, {Koptelova}, {Konstantinova}, {Kovalev}, {Kovalev}, {Kurtanidze}, {Kuss}, {Lande}, {Larionov}, {Latronico}, {Leto}, {Lindfors}, {Longo}, {Loparco}, {Lott}, {Lovellette}, {Lubrano}, {Madejski}, {Makeev}, {Marchegiani}, {Marscher}, {Marshall}, {Max-Moerbeck}, {Mazziotta}, {McConville}, {McEnery}, {Meurer}, {Michelson}, {Mitthumsiri}, {Mizuno}, {Moiseev}, {Monte}, {Monzani}, {Morselli}, {Moskalenko}, {Murgia}, {Nestoras}, {Nilsson}, {Nizhelsky}, {Nolan}, {Norris}, {Nuss}, {Ohsugi}, {Ojha}, {Omodei}, {Orlando}, {Ormes}, {Osborne}, {Ozaki}, {Pacciani}, {Padovani},
  {Pagani}, {Page}, {Paneque}, {Panetta}, {Parent}, {Pasanen}, {Pavlidou}, {Pelassa}, {Pepe}, {Perri}, {Pesce-Rollins}, {Piranomonte}, {Piron}, {Pittori}, {Porter}, {Puccetti}, {Rahoui}, {Rain{\`o}}, {Raiteri}, {Rando}, {Razzano}, {Reimer}, {Reimer}, {Reposeur}, {Richards}, {Ritz}, {Rochester}, {Rodriguez}, {Romani}, {Ros}, {Roth}, {Roustazadeh}, {Ryde}, {Sadrozinski}, {Sadun}, {Sanchez}, {Sander}, {Saz Parkinson}, {Scargle}, {Sellerholm}, {Sgr{\`o}}, {Shaw}, {Sigua}, {Siskind}, {Smith}, {Smith}, {Spandre}, {Spinelli}, {Starck}, {Stevenson}, {Stratta}, {Strickman}, {Suson}, {Tajima}, {Takahashi}, {Takahashi}, {Takalo}, {Tanaka}, {Thayer}, {Thayer}, {Thompson}, {Tibaldo}, {Torres}, {Tosti}, {Tramacere}, {Uchiyama}, {Usher}, {Vasileiou}, {Verrecchia}, {Vilchez}, {Villata}, {Vitale}, {Waite}, {Wang}, {Winer}, {Wood}, {Ylinen}, {Zensus}, {Zhekanis}, \& {Ziegler}}]{abdo2010}
{Abdo}, A.~A., {Ackermann}, M., {Agudo}, I., {et~al.} 2010, \apj, 716, 30, \dodoi{10.1088/0004-637X/716/1/30}

\bibitem[{{Abdollahi} {et~al.}(2020){Abdollahi}, {Acero}, {Ackermann}, {Ajello}, {Atwood}, {Axelsson}, {Baldini}, {Ballet}, {Barbiellini}, {Bastieri}, {Becerra Gonzalez}, {Bellazzini}, {Berretta}, {Bissaldi}, {Bland ford}, {Bloom}, {Bonino}, {Bottacini}, {Brandt}, {Bregeon}, {Bruel}, {Buehler}, {Burnett}, {Buson}, {Cameron}, {Caputo}, {Caraveo}, {Casandjian}, {Castro}, {Cavazzuti}, {Charles}, {Chaty}, {Chen}, {Cheung}, {Chiaro}, {Ciprini}, {Cohen-Tanugi}, {Cominsky}, {Coronado-Bl{\'a}zquez}, {Costantin}, {Cuoco}, {Cutini}, {D'Ammando}, {DeKlotz}, {de la Torre Luque}, {de Palma}, {Desai}, {Digel}, {Di Lalla}, {Di Mauro}, {Di Venere}, {Dom{\'\i}nguez}, {Dumora}, {Fana Dirirsa}, {Fegan}, {Ferrara}, {Franckowiak}, {Fukazawa}, {Funk}, {Fusco}, {Gargano}, {Gasparrini}, {Giglietto}, {Giommi}, {Giordano}, {Giroletti}, {Glanzman}, {Green}, {Grenier}, {Griffin}, {Grondin}, {Grove}, {Guiriec}, {Harding}, {Hayashi}, {Hays}, {Hewitt}, {Horan}, {J{\'o}hannesson}, {Johnson}, {Kamae}, {Kerr}, {Kocevski}, {Kovac'evic'},
  {Kuss}, {Landriu}, {Larsson}, {Latronico}, {Lemoine-Goumard}, {Li}, {Liodakis}, {Longo}, {Loparco}, {Lott}, {Lovellette}, {Lubrano}, {Madejski}, {Maldera}, {Malyshev}, {Manfreda}, {Marchesini}, {Marcotulli}, {Mart{\'\i}-Devesa}, {Martin}, {Massaro}, {Mazziotta}, {McEnery}, {Mereu}, {Meyer}, {Michelson}, {Mirabal}, {Mizuno}, {Monzani}, {Morselli}, {Moskalenko}, {Negro}, {Nuss}, {Ojha}, {Omodei}, {Orienti}, {Orlando}, {Ormes}, {Palatiello}, {Paliya}, {Paneque}, {Pei}, {Pe{\~n}a-Herazo}, {Perkins}, {Persic}, {Pesce-Rollins}, {Petrosian}, {Petrov}, {Piron}, {Poon}, {Porter}, {Principe}, {Rain{\`o}}, {Rando}, {Razzano}, {Razzaque}, {Reimer}, {Reimer}, {Remy}, {Reposeur}, {Romani}, {Saz Parkinson}, {Schinzel}, {Serini}, {Sgr{\`o}}, {Siskind}, {Smith}, {Spandre}, {Spinelli}, {Strong}, {Suson}, {Tajima}, {Takahashi}, {Tak}, {Thayer}, {Thompson}, {Tibaldo}, {Torres}, {Torresi}, {Valverde}, {Van Klaveren}, {van Zyl}, {Wood}, {Yassine}, \& {Zaharijas}}]{abdollahi2020}
{Abdollahi}, S., {Acero}, F., {Ackermann}, M., {et~al.} 2020, \apjs, 247, 33, \dodoi{10.3847/1538-4365/ab6bcb}

\bibitem[{{Abdollahi} {et~al.}(2023){Abdollahi}, {Ajello}, {Baldini}, {Ballet}, {Bastieri}, {Becerra Gonzalez}, {Bellazzini}, {Berretta}, {Bissaldi}, {Bonino}, {Brill}, {Bruel}, {Burns}, {Buson}, {Cameron}, {Caputo}, {Caraveo}, {Cibrario}, {Ciprini}, {Cristarella Orestano}, {Crnogorcevic}, {Cutini}, {D'Ammando}, {De Gaetano}, {Digel}, {Di Lalla}, {Di Venere}, {Dom{\'\i}nguez}, {Ramazani}, {Fegan}, {Ferrara}, {Fiori}, {Fleischhack}, {Franckowiak}, {Fukazawa}, {Fusco}, {Gammaldi}, {Gargano}, {Garrappa}, {Gasbarra}, {Gasparrini}, {Giglietto}, {Giordano}, {Giroletti}, {Green}, {Grenier}, {Guiriec}, {Gustafsson}, {Hays}, {Horan}, {Hou}, {J{\'o}hannesson}, {Kerr}, {Kocevski}, {Kuss}, {Latronico}, {Li}, {Liodakis}, {Longo}, {Loparco}, {Lorusso}, {Lott}, {Lovellette}, {Lubrano}, {Maldera}, {Manfreda}, {Mart{\'\i}-Devesa}, {Mazziotta}, {Mereu}, {Meyer}, {Michelson}, {Mizuno}, {Monzani}, {Morselli}, {Moskalenko}, {Negro}, {Omodei}, {Orlando}, {Ormes}, {Paneque}, {Panzarini}, {Perkins}, {Persic}, {Pesce-Rollins},
  {Pillera}, {Porter}, {Principe}, {Racusin}, {Rain{\`o}}, {Rando}, {Rani}, {Razzano}, {Razzaque}, {Reimer}, {Reimer}, {S{\'a}nchez-Conde}, {Parkinson}, {Scargle}, {Scotton}, {Serini}, {Sgr{\`o}}, {Siskind}, {Spandre}, {Spinelli}, {Suson}, {Tajima}, {Thompson}, {Torres}, {Valverde}, {Venters}, {Wadiasingh}, {Wagner}, \& {Wood}}]{abdollahi2023}
{Abdollahi}, S., {Ajello}, M., {Baldini}, L., {et~al.} 2023, \apjs, 265, 31, \dodoi{10.3847/1538-4365/acbb6a}

\bibitem[{{Agudo} {et~al.}(2012){Agudo}, {Molina}, {G{\'o}mez}, {Marscher}, {Jorstad}, \& {Heidt}}]{agudo2012}
{Agudo}, I., {Molina}, S.~N., {G{\'o}mez}, J.~L., {et~al.} 2012, in International Journal of Modern Physics Conference Series, Vol.~8, International Journal of Modern Physics Conference Series, 299--302, \dodoi{10.1142/S2010194512004746}

\bibitem[{{Agudo} {et~al.}(2018{\natexlab{a}}){Agudo}, {Thum}, {Ramakrishnan}, {Molina}, {Casadio}, \& {G{\'o}mez}}]{Agudo2018b}
{Agudo}, I., {Thum}, C., {Ramakrishnan}, V., {et~al.} 2018{\natexlab{a}}, \mnras, 473, 1850, \dodoi{10.1093/mnras/stx2437}

\bibitem[{{Agudo} {et~al.}(2018{\natexlab{b}}){Agudo}, {Thum}, {Molina}, {Casadio}, {Wiesemeyer}, {Morris}, {Paubert}, {G{\'o}mez}, \& {Kramer}}]{agudo2018}
{Agudo}, I., {Thum}, C., {Molina}, S.~N., {et~al.} 2018{\natexlab{b}}, \mnras, 474, 1427, \dodoi{10.1093/mnras/stx2435}

\bibitem[{{Aharonian}(2000)}]{aharonian2000}
{Aharonian}, F.~A. 2000, New Astronomy, 5, 377, \dodoi{10.1016/S1384-1076(00)00039-7}

\bibitem[{{Ajello} {et~al.}(2020){Ajello}, {Angioni}, {Axelsson}, {Ballet}, {Barbiellini}, {Bastieri}, {Becerra Gonzalez}, {Bellazzini}, {Bissaldi}, {Bloom}, {Bonino}, {Bottacini}, {Bruel}, {Buson}, {Cafardo}, {Cameron}, {Cavazzuti}, {Chen}, {Cheung}, {Ciprini}, {Costantin}, {Cutini}, {D'Ammando}, {de la Torre Luque}, {de Menezes}, {de Palma}, {Desai}, {Di Lalla}, {Di Venere}, {Dom{\'\i}nguez}, {Dirirsa}, {Ferrara}, {Finke}, {Franckowiak}, {Fukazawa}, {Funk}, {Fusco}, {Gargano}, {Garrappa}, {Gasparrini}, {Giglietto}, {Giordano}, {Giroletti}, {Green}, {Grenier}, {Guiriec}, {Harita}, {Hays}, {Horan}, {Itoh}, {J{\'o}hannesson}, {Kovac'evic'}, {Krauss}, {Kreter}, {Kuss}, {Larsson}, {Leto}, {Li}, {Liodakis}, {Longo}, {Loparco}, {Lott}, {Lovellette}, {Lubrano}, {Madejski}, {Maldera}, {Manfreda}, {Mart{\'\i}-Devesa}, {Massaro}, {Mazziotta}, {Mereu}, {Meyer}, {Migliori}, {Mirabal}, {Mizuno}, {Monzani}, {Morselli}, {Moskalenko}, {Negro}, {Nemmen}, {Nuss}, {Ojha}, {Ojha}, {Omodei}, {Orienti}, {Orlando}, {Ormes},
  {Paliya}, {Pei}, {Pe{\~n}a-Herazo}, {Persic}, {Pesce-Rollins}, {Petrov}, {Piron}, {Poon}, {Principe}, {Rain{\`o}}, {Rando}, {Rani}, {Razzano}, {Razzaque}, {Reimer}, {Reimer}, {Schinzel}, {Serini}, {Sgr{\`o}}, {Siskind}, {Spandre}, {Spinelli}, {Suson}, {Tachibana}, {Thompson}, {Torres}, {Torresi}, {Troja}, {Valverde}, {van Zyl}, \& {Yassine}}]{ajello2020}
{Ajello}, M., {Angioni}, R., {Axelsson}, M., {et~al.} 2020, \apj, 892, 105, \dodoi{10.3847/1538-4357/ab791e}

\bibitem[{{Atwood} {et~al.}(2009){Atwood}, {Abdo}, {Ackermann}, {Althouse}, {Anderson}, {Axelsson}, {Baldini}, {Ballet}, {Band}, {Barbiellini}, {Bartelt}, {Bastieri}, {Baughman}, {Bechtol}, {B{\'e}d{\'e}r{\`e}de}, {Bellardi}, {Bellazzini}, {Berenji}, {Bignami}, {Bisello}, {Bissaldi}, {Blandford}, {Bloom}, {Bogart}, {Bonamente}, {Bonnell}, {Borgland }, {Bouvier}, {Bregeon}, {Brez}, {Brigida}, {Bruel}, {Burnett}, {Busetto}, {Caliandro}, {Cameron}, {Caraveo}, {Carius}, {Carlson}, {Casandjian}, {Cavazzuti}, {Ceccanti}, {Cecchi}, {Charles}, {Chekhtman}, {Cheung}, {Chiang}, {Chipaux}, {Cillis}, {Ciprini}, {Claus}, {Cohen-Tanugi}, {Condamoor}, {Conrad}, {Corbet}, {Corucci}, {Costamante}, {Cutini}, {Davis}, {Decotigny}, {DeKlotz}, {Dermer}, {de Angelis}, {Digel}, {do Couto e Silva}, {Drell}, {Dubois}, {Dumora}, {Edmonds}, {Fabiani}, {Farnier}, {Favuzzi}, {Flath}, {Fleury}, {Focke}, {Funk}, {Fusco}, {Gargano}, {Gasparrini}, {Gehrels}, {Gentit}, {Germani}, {Giebels}, {Giglietto}, {Giommi}, {Giordano}, {Glanzman},
  {Godfrey}, {Grenier}, {Grondin}, {Grove}, {Guillemot}, {Guiriec}, {Haller}, {Harding}, {Hart}, {Hays}, {Healey}, {Hirayama}, {Hjalmarsdotter}, {Horn}, {Hughes}, {J{\'o}hannesson}, {Johansson}, {Johnson}, {Johnson}, {Johnson}, {Johnson}, {Kamae}, {Katagiri}, {Kataoka}, {Kavelaars}, {Kawai}, {Kelly}, {Kerr}, {Klamra}, {Kn{\"o}dlseder}, {Kocian}, {Komin}, {Kuehn}, {Kuss}, {Landriu}, {Latronico}, {Lee}, {Lee}, {Lemoine-Goumard}, {Lionetto}, {Longo}, {Loparco}, {Lott}, {Lovellette}, {Lubrano}, {Madejski}, {Makeev}, {Marangelli}, {Massai}, {Mazziotta}, {McEnery}, {Menon}, {Meurer}, {Michelson}, {Minuti}, {Mirizzi}, {Mitthumsiri}, {Mizuno}, {Moiseev}, {Monte}, {Monzani}, {Moretti}, {Morselli}, {Moskalenko}, {Murgia}, {Nakamori}, {Nishino}, {Nolan}, {Norris}, {Nuss}, {Ohno}, {Ohsugi}, {Omodei}, {Orlando}, {Ormes}, {Paccagnella}, {Paneque}, {Panetta}, {Parent}, {Pearce}, {Pepe}, {Perazzo}, {Pesce-Rollins}, {Picozza}, {Pieri}, {Pinchera}, {Piron}, {Porter}, {Poupard}, {Rain{\`o}}, {Rando}, {Rapposelli}, {Razzano},
  {Reimer}, {Reimer}, {Reposeur}, {Reyes}, {Ritz}, {Rochester}, {Rodriguez}, {Romani}, {Roth}, {Russell}, {Ryde}, {Sabatini}, {Sadrozinski}, {Sanchez}, {Sand er}, {Sapozhnikov}, {Parkinson}, {Scargle}, {Schalk}, {Scolieri}, {Sgr{\`o}}, {Share}, {Shaw}, {Shimokawabe}, {Shrader}, {Sierpowska-Bartosik}, {Siskind}, {Smith}, {Smith}, {Spandre}, {Spinelli}, {Starck}, {Stephens}, {Strickman}, {Strong}, {Suson}, {Tajima}, {Takahashi}, {Takahashi}, {Tanaka}, {Tenze}, {Tether}, {Thayer}, {Thayer}, {Thompson}, {Tibaldo}, {Tibolla}, {Torres}, {Tosti}, {Tramacere}, {Turri}, {Usher}, {Vilchez}, {Vitale}, {Wang}, {Watters}, {Winer}, {Wood}, {Ylinen}, \& {Ziegler}}]{atwood2009}
{Atwood}, W.~B., {Abdo}, A.~A., {Ackermann}, M., {et~al.} 2009, \apj, 697, 1071, \dodoi{10.1088/0004-637X/697/2/1071}

\bibitem[{{Bachev} {et~al.}(2023){Bachev}, {Tripathi}, {Gupta}, {Kushwaha}, {Strigachev}, {Kurtenkov}, {Nikolov}, {Boeva}, {Damljanovic}, {Vince}, {Stojanovic}, {Kishore}, {Gaur}, {Dhiman}, {Fan}, {Kalita}, {Spassov}, \& {Semkov}}]{Bachev2023}
{Bachev}, R., {Tripathi}, T., {Gupta}, A.~C., {et~al.} 2023, \mnras, 522, 3018, \dodoi{10.1093/mnras/stad1063}

\bibitem[{{Baldini} {et~al.}(2022){Baldini}, {Bucciantini}, {Lalla}, {Ehlert}, {Manfreda}, {Negro}, {Omodei}, {Pesce-Rollins}, {Sgr{\`o}}, \& {Silvestri}}]{baldini2022}
{Baldini}, L., {Bucciantini}, N., {Lalla}, N.~D., {et~al.} 2022, SoftwareX, 19, 101194, \dodoi{10.1016/j.softx.2022.101194}

\bibitem[{{Ballet} {et~al.}(2020){Ballet}, {Burnett}, {Digel}, \& {Lott}}]{ballet2020}
{Ballet}, J., {Burnett}, T.~H., {Digel}, S.~W., \& {Lott}, B. 2020, arXiv e-prints, arXiv:2005.11208, \dodoi{10.48550/arXiv.2005.11208}

\bibitem[{{Bania} {et~al.}(1991){Bania}, {Marscher}, \& {Barvainis}}]{Bania1991}
{Bania}, T.~M., {Marscher}, A.~P., \& {Barvainis}, R. 1991, \aj, 101, 2147, \dodoi{10.1086/115836}

\bibitem[{{Barniol Duran} {et~al.}(2017){Barniol Duran}, {Tchekhovskoy}, \& {Giannios}}]{Barniol2017}
{Barniol Duran}, R., {Tchekhovskoy}, A., \& {Giannios}, D. 2017, \mnras, 469, 4957, \dodoi{10.1093/mnras/stx1165}

\bibitem[{{Begelman} \& {Sikora}(1987)}]{Begelman1987}
{Begelman}, M.~C., \& {Sikora}, M. 1987, \apj, 322, 650, \dodoi{10.1086/165760}

\bibitem[{{Blackburn}(1995)}]{1995ASPC...77..367B}
{Blackburn}, J.~K. 1995, in Astronomical Society of the Pacific Conference Series, Vol.~77, Astronomical Data Analysis Software and Systems IV, ed. R.~A. {Shaw}, H.~E. {Payne}, \& J.~J.~E. {Hayes}, 367

\bibitem[{{Blandford} {et~al.}(2019){Blandford}, {Meier}, \& {Readhead}}]{blandford2019}
{Blandford}, R., {Meier}, D., \& {Readhead}, A. 2019, \araa, 57, 467, \dodoi{10.1146/annurev-astro-081817-051948}

\bibitem[{{Blinov} {et~al.}(2018){Blinov}, {Pavlidou}, {Papadakis}, {Kiehlmann}, {Liodakis}, {Panopoulou}, {Angelakis}, {Balokovi{\'c}}, {Hovatta}, {King}, {Kus}, {Kylafis}, {Mahabal}, {Maharana}, {Myserlis}, {Paleologou}, {Papamastorakis}, {Pazderski}, {Pearson}, {Ramaprakash}, {Readhead}, {Reig}, {Tassis}, \& {Zensus}}]{Blinov2018}
{Blinov}, D., {Pavlidou}, V., {Papadakis}, I., {et~al.} 2018, \mnras, 474, 1296, \dodoi{10.1093/mnras/stx2786}

\bibitem[{{Blinov} {et~al.}(2021){Blinov}, {Kiehlmann}, {Pavlidou}, {Panopoulou}, {Skalidis}, {Angelakis}, {Casadio}, {Einoder}, {Hovatta}, {Kokolakis}, {Kougentakis}, {Kus}, {Kylafis}, {Kyritsis}, {Lalakos}, {Liodakis}, {Maharana}, {Makrydopoulou}, {Mandarakas}, {Maragkakis}, {Myserlis}, {Papadakis}, {Paterakis}, {Pearson}, {Ramaprakash}, {Readhead}, {Reig}, {S{\l}owikowska}, {Tassis}, {Xexakis}, {{\.Z}ejmo}, \& {Zensus}}]{Blinov2021}
{Blinov}, D., {Kiehlmann}, S., {Pavlidou}, V., {et~al.} 2021, \mnras, 501, 3715, \dodoi{10.1093/mnras/staa3777}

\bibitem[{{Bonometto} \& {Saggion}(1973)}]{Bonometto1973}
{Bonometto}, S., \& {Saggion}, A. 1973, \aap, 23, 9

\bibitem[{{Burrows} {et~al.}(2005){Burrows}, {Hill}, {Nousek}, {Kennea}, {Wells}, {Osborne}, {Abbey}, {Beardmore}, {Mukerjee}, {Short}, {Chincarini}, {Campana}, {Citterio}, {Moretti}, {Pagani}, {Tagliaferri}, {Giommi}, {Capalbi}, {Tamburelli}, {Angelini}, {Cusumano}, {Br{\"a}uninger}, {Burkert}, \& {Hartner}}]{Bur05}
{Burrows}, D.~N., {Hill}, J.~E., {Nousek}, J.~A., {et~al.} 2005, \ssr, 120, 165, \dodoi{10.1007/s11214-005-5097-2}

\bibitem[{{Dermer} \& {Schlickeiser}(1993)}]{dermer1993}
{Dermer}, C.~D., \& {Schlickeiser}, R. 1993, \apj, 416, 458, \dodoi{10.1086/173251}

\bibitem[{{Di Gesu} {et~al.}(2022){Di Gesu}, {Tavecchio}, {Donnarumma}, {Marscher}, {Pesce-Rollins}, \& {Landoni}}]{digesu2022}
{Di Gesu}, L., {Tavecchio}, F., {Donnarumma}, I., {et~al.} 2022, \aas, 662, A83, \dodoi{10.1051/0004-6361/202243168}

\bibitem[{{Di Gesu} {et~al.}(2023){Di Gesu}, {Marshall}, {Ehlert}, {Kim}, {Donnarumma}, {Tavecchio}, {Liodakis}, {Kiehlmann}, {Agudo}, {Jorstad}, {Muleri}, {Marscher}, {Puccetti}, {Middei}, {Perri}, {Pacciani}, {Negro}, {Romani}, {Di Marco}, {Blinov}, {Bourbah}, {Kontopodis}, {Mandarakas}, {Romanopoulos}, {Skalidis}, {Vervelaki}, {Casadio}, {Escudero}, {Myserlis}, {Gurwell}, {Rao}, {Keating}, {Kouch}, {Lindfors}, {Aceituno}, {Bernardos}, {Bonnoli}, {Casanova}, {Garc{\'\i}a-Comas}, {Ag{\'\i}s-Gonz{\'a}lez}, {Husillos}, {Marchini}, {Sota}, {Imazawa}, {Sasada}, {Fukazawa}, {Kawabata}, {Uemura}, {Mizuno}, {Nakaoka}, {Akitaya}, {Savchenko}, {Vasilyev}, {G{\'o}mez}, {Antonelli}, {Barnouin}, {Bonino}, {Cavazzuti}, {Costamante}, {Chen}, {Cibrario}, {De Rosa}, {Di Pierro}, {Errando}, {Kaaret}, {Karas}, {Krawczynski}, {Lisalda}, {Madejski}, {Malacaria}, {Marin}, {Marinucci}, {Massaro}, {Matt}, {Mitsuishi}, {O'Dell}, {Paggi}, {Peirson}, {Petrucci}, {Ramsey}, {Tennant}, {Wu}, {Bachetti}, {Baldini}, {Baumgartner},
  {Bellazzini}, {Bianchi}, {Bongiorno}, {Brez}, {Bucciantini}, {Capitanio}, {Castellano}, {Ciprini}, {Costa}, {Del Monte}, {Di Lalla}, {Doroshenko}, {Dov{\v{c}}iak}, {Enoto}, {Evangelista}, {Fabiani}, {Ferrazzoli}, {Garcia}, {Gunji}, {Hayashida}, {Heyl}, {Iwakiri}, {Kislat}, {Kitaguchi}, {Kolodziejczak}, {La Monaca}, {Latronico}, {Maldera}, {Manfreda}, {Ng}, {Omodei}, {Oppedisano}, {Papitto}, {Pavlov}, {Pesce-Rollins}, {Pilia}, {Possenti}, {Poutanen}, {Rankin}, {Ratheesh}, {Roberts}, {Sgr{\`o}}, {Slane}, {Soffitta}, {Spandre}, {Swartz}, {Tamagawa}, {Taverna}, {Tawara}, {Thomas}, {Tombesi}, {Trois}, {Tsygankov}, {Turolla}, {Vink}, {Weisskopf}, {Xie}, \& {Zane}}]{DiGesu2023}
{Di Gesu}, L., {Marshall}, H.~L., {Ehlert}, S.~R., {et~al.} 2023, Nature Astronomy, 7, 1245, \dodoi{10.1038/s41550-023-02032-7}

\bibitem[{{Di Marco} {et~al.}(2022){Di Marco}, {Costa}, {Muleri}, {Soffitta}, {Fabiani}, {La Monaca}, {Rankin}, {Xie}, {Bachetti}, {Baldini}, {Baumgartner}, {Bellazzini}, {Brez}, {Castellano}, {Del Monte}, {Di Lalla}, {Ferrazzoli}, {Latronico}, {Maldera}, {Manfreda}, {O'Dell}, {Perri}, {Pesce-Rollins}, {Puccetti}, {Ramsey}, {Ratheesh}, {Sgr{\`o}}, {Spandre}, {Tennant}, {Tobia}, {Trois}, \& {Weisskopf}}]{dimarco2022}
{Di Marco}, A., {Costa}, E., {Muleri}, F., {et~al.} 2022, \aj, 163, 170, \dodoi{10.3847/1538-3881/ac51c9}

\bibitem[{{Di Marco} {et~al.}(2023){Di Marco}, {Soffitta}, {Costa}, {Ferrazzoli}, {La Monaca}, {Rankin}, {Ratheesh}, {Xie}, {Baldini}, {Del Monte}, {Ehlert}, {Fabiani}, {Kim}, {Muleri}, {O'Dell}, {Ramsey}, {Rubini}, {Sgr{\`o}}, {Silvestri}, {Tennant}, \& {Weisskopf}}]{DiMarco2023}
{Di Marco}, A., {Soffitta}, P., {Costa}, E., {et~al.} 2023, \aj, 165, 143, \dodoi{10.3847/1538-3881/acba0f}

\bibitem[{{Escudero Pedrosa} {et~al.}(2024{\natexlab{a}}){Escudero Pedrosa}, {Morcuende Parrilla}, \& {Otero-Santos}}]{escudero2023}
{Escudero Pedrosa}, J., {Morcuende Parrilla}, D., \& {Otero-Santos}, J. 2024{\natexlab{a}}, {IOP4}, v1.2.0,  Zenodo, \dodoi{10.5281/zenodo.10222722}

\bibitem[{{Escudero Pedrosa} {et~al.}(2024{\natexlab{b}}){Escudero Pedrosa}, {Agudo}, {Morcuende}, {Otero-Santos}, {Bonnoli}, {Piirola}, {Husillos}, {Bernardos}, {L{\'o}pez-Coto}, {Sota}, {Casanova}, {Aceituno}, \& {Santos-Sanz}}]{escudero2024}
{Escudero Pedrosa}, J., {Agudo}, I., {Morcuende}, D., {et~al.} 2024{\natexlab{b}}, \aj, 168, 84, \dodoi{10.3847/1538-3881/ad5a80}

\bibitem[{{Gabriel} {et~al.}(2004){Gabriel}, {Denby}, {Fyfe}, {Hoar}, {Ibarra}, {Ojero}, {Osborne}, {Saxton}, {Lammers}, \& {Vacanti}}]{Gabriel2004}
{Gabriel}, C., {Denby}, M., {Fyfe}, D.~J., {et~al.} 2004, in Astronomical Society of the Pacific Conference Series, Vol. 314, Astronomical Data Analysis Software and Systems (ADASS) XIII, ed. F.~{Ochsenbein}, M.~G. {Allen}, \& D.~{Egret}, 759

\bibitem[{{Gehrels} {et~al.}(2004){Gehrels}, {Chincarini}, {Giommi}, \& et~al.}]{Gehrels2004}
{Gehrels}, N., {Chincarini}, G., {Giommi}, P., \& et~al. 2004, \apj, 611, 38, \dodoi{10.1086/422091}

\bibitem[{{Guan} {et~al.}(2014){Guan}, {Li}, \& {Li}}]{Guan2014}
{Guan}, X., {Li}, H., \& {Li}, S. 2014, \apj, 781, 48, \dodoi{10.1088/0004-637X/781/1/48}

\bibitem[{{Gurwell} {et~al.}(2007){Gurwell}, {Peck}, {Hostler}, {Darrah}, \& {Katz}}]{2007ASPC..375..234G}
{Gurwell}, M.~A., {Peck}, A.~B., {Hostler}, S.~R., {Darrah}, M.~R., \& {Katz}, C.~A. 2007, in Astronomical Society of the Pacific Conference Series, Vol. 375, From Z-Machines to ALMA: (Sub)Millimeter Spectroscopy of Galaxies, ed. A.~J. {Baker}, J.~{Glenn}, A.~I. {Harris}, J.~G. {Mangum}, \& M.~S. {Yun}, 234

\bibitem[{{Harrison} {et~al.}(2013){Harrison}, {Craig}, {Christensen}, {Hailey}, {Zhang}, {Boggs}, {Stern}, {Cook}, {Forster}, {Giommi}, {Grefenstette}, {Kim}, {Kitaguchi}, {Koglin}, {Madsen}, {Mao}, {Miyasaka}, {Mori}, {Perri}, {Pivovaroff}, {Puccetti}, {Rana}, {Westergaard}, {Willis}, {Zoglauer}, {An}, {Bachetti}, {Barri{\`e}re}, {Bellm}, {Bhalerao}, {Brejnholt}, {Fuerst}, {Liebe}, {Markwardt}, {Nynka}, {Vogel}, {Walton}, {Wik}, {Alexander}, {Cominsky}, {Hornschemeier}, {Hornstrup}, {Kaspi}, {Madejski}, {Matt}, {Molendi}, {Smith}, {Tomsick}, {Ajello}, {Ballantyne}, {Balokovi{\'c}}, {Barret}, {Bauer}, {Blandford}, {Brandt}, {Brenneman}, {Chiang}, {Chakrabarty}, {Chenevez}, {Comastri}, {Dufour}, {Elvis}, {Fabian}, {Farrah}, {Fryer}, {Gotthelf}, {Grindlay}, {Helfand}, {Krivonos}, {Meier}, {Miller}, {Natalucci}, {Ogle}, {Ofek}, {Ptak}, {Reynolds}, {Rigby}, {Tagliaferri}, {Thorsett}, {Treister}, \& {Urry}}]{Harrison2013}
{Harrison}, F.~A., {Craig}, W.~W., {Christensen}, F.~E., {et~al.} 2013, \apj, 770, 103, \dodoi{10.1088/0004-637X/770/2/103}

\bibitem[{{Ho} {et~al.}(2004){Ho}, {Moran}, \& {Lo}}]{Ho2004}
{Ho}, P. T.~P., {Moran}, J.~M., \& {Lo}, K.~Y. 2004, \apjl, 616, L1, \dodoi{10.1086/423245}

\bibitem[{{Hovatta} \& {Lindfors}(2019)}]{Hovatta2019}
{Hovatta}, T., \& {Lindfors}, E. 2019, \nar, 87, 101541, \dodoi{10.1016/j.newar.2020.101541}

\bibitem[{{Hovatta} {et~al.}(2016){Hovatta}, {Lindfors}, {Blinov}, {Pavlidou}, {Nilsson}, {Kiehlmann}, {Angelakis}, {Fallah Ramazani}, {Liodakis}, {Myserlis}, {Panopoulou}, \& {Pursimo}}]{Hovatta2016}
{Hovatta}, T., {Lindfors}, E., {Blinov}, D., {et~al.} 2016, \aap, 596, A78, \dodoi{10.1051/0004-6361/201628974}

\bibitem[{{IceCube Collaboration} {et~al.}(2018){IceCube Collaboration}, {Aartsen}, {Ackermann}, {Adams}, {Aguilar}, {Ahlers}, {Ahrens}, {Samarai}, {Altmann}, {Andeen}, {Anderson}, {Ansseau}, {Anton}, {Arg{\"u}elles}, {Arsioli}, {Auffenberg}, {Axani}, {Bagherpour}, {Bai}, {Barron}, {Barwick}, {Baum}, {Bay}, {Beatty}, {Becker Tjus}, {Becker}, {BenZvi}, {Berley}, {Bernardini}, {Besson}, {Binder}, {Bindig}, {Blaufuss}, {Blot}, {Bohm}, {B{\"o}rner}, {Bos}, {B{\"o}ser}, {Botner}, {Bourbeau}, {Bourbeau}, {Bradascio}, {Braun}, {Brenzke}, {Bretz}, {Bron}, {Brostean-Kaiser}, {Burgman}, {Busse}, {Carver}, {Cheung}, {Chirkin}, {Christov}, {Clark}, {Classen}, {Coenders}, {Collin}, {Conrad}, {Coppin}, {Correa}, {Cowen}, {Cross}, {Dave}, {Day}, {de Andr{\'e}}, {De Clercq}, {DeLaunay}, {Dembinski}, {DeRidder}, {Desiati}, {de Vries}, {de Wasseige}, {de With}, {DeYoung}, {D{\'\i}az-V{\'e}lez}, {di Lorenzo}, {Dujmovic}, {Dumm}, {Dunkman}, {Dvorak}, {Eberhardt}, {Ehrhardt}, {Eichmann}, {Eller}, {Evenson}, {Fahey}, {Fazely},
  {Felde}, {Filimonov}, {Finley}, {Flis}, {Franckowiak}, {Friedman}, {Fritz}, {Gaisser}, {Gallagher}, {Gerhardt}, {Ghorbani}, {Giommi}, {Glauch}, {Gl{\"u}senkamp}, {Goldschmidt}, {Gonzalez}, {Grant}, {Griffith}, {Haack}, {Hallgren}, {Halzen}, {Hanson}, {Hebecker}, {Heereman}, {Helbing}, {Hellauer}, {Hickford}, {Hignight}, {Hill}, {Hoffman}, {Hoffmann}, {Hoinka}, {Hokanson-Fasig}, {Hoshina}, {Huang}, {Huber}, {Hultqvist}, {H{\"u}nnefeld}, {Hussain}, {In}, {Iovine}, {Ishihara}, {Jacobi}, {Japaridze}, {Jeong}, {Jero}, {Jones}, {Kalaczynski}, {Kang}, {Kappes}, {Kappesser}, {Karg}, {Karle}, {Katz}, {Kauer}, {Keivani}, {Kelley}, {Kheirandish}, {Kim}, {Kim}, {Kintscher}, {Kiryluk}, {Kittler}, {Klein}, {Koirala}, {Kolanoski}, {K{\"o}pke}, {Kopper}, {Kopper}, {Koschinsky}, {Koskinen}, {Kowalski}, {Krammer}, {Krings}, {Kroll}, {Kr{\"u}ckl}, {Kunwar}, {Kurahashi}, {Kuwabara}, {Kyriacou}, {Labare}, {Lanfranchi}, {Larson}, {Lauber}, {Leonard}, {Lesiak-Bzdak}, {Leuermann}, {Liu}, {Lozano Mariscal}, {Lu}, {L{\"u}nemann},
  {Luszczak}, {Madsen}, {Maggi}, {Mahn}, {Mancina}, {Maruyama}, {Mase}, {Maunu}, {Meagher}, {Medici}, {Meier}, {Menne}, {Merino}, {Meures}, {Miarecki}, {Micallef}, {Moment{\'e}}, {Montaruli}, {Moore}, {Morse}, {Moulai}, {Nahnhauer}, {Nakarmi}, {Naumann}, {Neer}, {Niederhausen}, {Nowicki}, {Nygren}, {Obertacke Pollmann}, {Olivas}, {O'Murchadha}, {O'Sullivan}, {Padovani}, {Palczewski}, {Pandya}, {Pankova}, {Peiffer}, {Pepper}, {P{\'e}rez de los Heros}, {Pieloth}, {Pinat}, {Plum}, {Price}, {Przybylski}, {Raab}, {R{\"a}del}, {Rameez}, {Rawlins}, {Rea}, {Reimann}, {Relethford}, {Relich}, {Resconi}, {Rhode}, {Richman}, {Robertson}, {Rongen}, {Rott}, {Ruhe}, {Ryckbosch}, {Rysewyk}, {Safa}, {Sahakyan}, {S{\"a}lzer}, {Sanchez Herrera}, {Sandrock}, {Sandroos}, {Santander}, {Sarkar}, {Sarkar}, {Satalecka}, {Schlunder}, {Schmidt}, {Schneider}, {Schoenen}, {Sch{\"o}neberg}, {Schumacher}, {Sclafani}, {Seckel}, {Seunarine}, {Soedingrekso}, {Soldin}, {Song}, {Spiczak}, {Spiering}, {Stachurska}, {Stamatikos}, {Stanev},
  {Stasik}, {Stettner}, {Steuer}, {Stezelberger}, {Stokstad}, {St{\"o}{\ss}l}, {Strotjohann}, {Stuttard}, {Sullivan}, {Sutherland}, {Taboada}, {Tatar}, {Tenholt}, {Ter-Antonyan}, {Terliuk}, {Tilav}, {Toale}, {Tobin}, {Toennis}, {Toscano}, {Tosi}, {Tselengidou}, {Tung}, {Turcati}, {Turley}, {Ty}, {Unger}, {Usner}, {Vandenbroucke}, {Van Driessche}, {van Eijk}, {van Eijndhoven}, {Vanheule}, {van Santen}, {Vogel}, {Vraeghe}, {Walck}, {Wallace}, {Wallraff}, {Wandler}, {Wandkowsky}, {Waza}, {Weaver}, {Weiss}, {Wendt}, {Werthebach}, {Westerhoff}, {Whelan}, {Whitehorn}, {Wiebe}, {Wiebusch}, {Wille}, {Williams}, {Wills}, {Wolf}, {Wood}, {Wood}, {Woschnagg}, {Xu}, {Xu}, {Xu}, {Yanez}, {Yodh}, {Yoshida}, \& {Yuan}}]{Icecube2018}
{IceCube Collaboration}, {Aartsen}, M.~G., {Ackermann}, M., {et~al.} 2018, Science, 361, 147, \dodoi{10.1126/science.aat2890}

\bibitem[{{Jansen} {et~al.}(2001){Jansen}, {Lumb}, {Altieri}, {Clavel}, {Ehle}, {Erd}, {Gabriel}, {Guainazzi}, {Gondoin}, {Much}, {Munoz}, {Santos}, {Schartel}, {Texier}, \& {Vacanti}}]{Jansen2001}
{Jansen}, F., {Lumb}, D., {Altieri}, B., {et~al.} 2001, \aap, 365, L1, \dodoi{10.1051/0004-6361:20000036}

\bibitem[{{Jethwa} {et~al.}(2015){Jethwa}, {Saxton}, {Guainazzi}, {Rodriguez-Pascual}, \& {Stuhlinger}}]{Jethwa2015}
{Jethwa}, P., {Saxton}, R., {Guainazzi}, M., {Rodriguez-Pascual}, P., \& {Stuhlinger}, M. 2015, \aap, 581, A104, \dodoi{10.1051/0004-6361/201425579}

\bibitem[{{Jorstad} {et~al.}(2010){Jorstad}, {Marscher}, {Larionov}, {Agudo}, {Smith}, {Gurwell}, {L{\"a}hteenm{\"a}ki}, {Tornikoski}, {Markowitz}, {Arkharov}, {Blinov}, {Chatterjee}, {D'Arcangelo}, {Falcone}, {G{\'o}mez}, {Hagen-Thorn}, {Jordan}, {Kimeridze}, {Konstantinova}, {Kopatskaya}, {Kurtanidze}, {Larionova}, {Larionova}, {McHardy}, {Melnichuk}, {Roca-Sogorb}, {Schmidt}, {Skiff}, {Taylor}, {Thum}, {Troitsky}, \& {Wiesemeyer}}]{Jorstad2010}
{Jorstad}, S.~G., {Marscher}, A.~P., {Larionov}, V.~M., {et~al.} 2010, \apj, 715, 362, \dodoi{10.1088/0004-637X/715/1/362}

\bibitem[{{Jorstad} {et~al.}(2017){Jorstad}, {Marscher}, {Morozova}, {Troitsky}, {Agudo}, {Casadio}, {Foord}, {G{\'o}mez}, {MacDonald}, {Molina}, {L{\"a}hteenm{\"a}ki}, {Tammi}, \& {Tornikoski}}]{Jorstad2017}
{Jorstad}, S.~G., {Marscher}, A.~P., {Morozova}, D.~A., {et~al.} 2017, \apj, 846, 98, \dodoi{10.3847/1538-4357/aa8407}

\bibitem[{{Kang} {et~al.}(2015){Kang}, {Lee}, \& {Byun}}]{Kang2015}
{Kang}, S., {Lee}, S.-S., \& {Byun}, D.-Y. 2015, Journal of Korean Astronomical Society, 48, 257, \dodoi{10.5303/JKAS.2015.48.5.257}

\bibitem[{{Kim} {et~al.}(2024){Kim}, {Di Gesu}, {Liodakis}, {Marscher}, {Jorstad}, {Middei}, {Marshall}, {Pacciani}, {Agudo}, {Tavecchio}, {Cibrario}, {Tugliani}, {Bonino}, {Negro}, {Puccetti}, {Tombesi}, {Costa}, {Donnarumma}, {Soffitta}, {Mizuno}, {Fukazawa}, {Kawabata}, {Nakaoka}, {Uemura}, {Imazawa}, {Sasada}, {Akitaya}, {Jos{\`e} Aceituno}, {Bonnoli}, {Casanova}, {Myserlis}, {Sievers}, {Angelakis}, {Kraus}, {Yeon Cheong}, {Jeong}, {Kang}, {Kim}, {Lee}, {Ag{\`\i}s-Gonz{\`a}lez}, {Sota}, {Escudero}, {Gurwell}, {Keating}, {Rao}, {Kouch}, {Lindfors}, {Bourbah}, {Kiehlmann}, {Kontopodis}, {Mandarakas}, {Romanopoulos}, {Skalidis}, {Vervelaki}, {Savchenko}, {Antonelli}, {Bachetti}, {Baldini}, {Baumgartner}, {Bellazzini}, {Bianchi}, {Bongiorno}, {Brez}, {Bucciantini}, {Capitanio}, {Castellano}, {Cavazzuti}, {Chen}, {Ciprini}, {De Rosa}, {Del Monte}, {Di Lalla}, {Di Marco}, {Doroshenko}, {Dov{\v{c}}iak}, {Ehlert}, {Enoto}, {Evangelista}, {Fabiani}, {Ferrazzoli}, {Garcia}, {Gunji}, {Hayashida}, {Heyl}, {Iwakiri},
  {Kaaret}, {Karas}, {Kislat}, {Kitaguchi}, {Kolodziejczak}, {Krawczynski}, {La Monaca}, {Latronico}, {Maldera}, {Manfreda}, {Marin}, {Marinucci}, {Massaro}, {Matt}, {Mitsuishi}, {Muleri}, {Ng}, {O'Dell}, {Omodei}, {Oppedisano}, {Papitto}, {Pavlov}, {Peirson}, {Perri}, {Pesce-Rollins}, {Petrucci}, {Pilia}, {Possenti}, {Poutanen}, {Ramsey}, {Rankin}, {Ratheesh}, {Roberts}, {Romani}, {Sgr{\'o}}, {Slane}, {Spandre}, {Swartz}, {Tamagawa}, {Taverna}, {Tawara}, {Tennant}, {Thomas}, {Trois}, {Tsygankov}, {Turolla}, {Vink}, {Weisskopf}, {Wu}, {Xie}, \& {Zane}}]{2024A&A...681A..12K}
{Kim}, D.~E., {Di Gesu}, L., {Liodakis}, I., {et~al.} 2024, \aap, 681, A12, \dodoi{10.1051/0004-6361/202347408}

\bibitem[{{Kislat} {et~al.}(2015){Kislat}, {Clark}, {Beilicke}, \& {Krawczynski}}]{Kislat2015}
{Kislat}, F., {Clark}, B., {Beilicke}, M., \& {Krawczynski}, H. 2015, Astroparticle Physics, 68, 45, \dodoi{10.1016/j.astropartphys.2015.02.007}

\bibitem[{{Kouch} {et~al.}(2024{\natexlab{a}}){Kouch}, {Liodakis}, {Fenu}, {Zhang}, {Boula}, {Middei}, {Di Gesu}, {Paraschos}, {Agudo}, {Jorstad}, {Lindfors}, {Marscher}, {Krawczynski}, {Negro}, {Hu}, {Kim}, {Cavazzuti}, {Errando}, {Blinov}, {Gourni}, {Kiehlmann}, {Kourtidis}, {Mandarakas}, {Triantafyllou}, {Vervelaki}, {Borman}, {Kopatskaya}, {Larionova}, {Morozova}, {Savchenko}, {Vasilyev}, {Troitskiy}, {Grishina}, {Zhovtan}, {Jos{\'e} Aceituno}, {Bonnoli}, {Casanova}, {Escudero}, {Ag{\'\i}s-Gonz{\'a}lez}, {Husillos}, {Otero-Santos}, {Piirola}, {Sota}, {Myserlis}, {Gurwell}, {Keating}, {Rao}, {Angelakis}, {Kraus}, {Antonelli}, {Bachetti}, {Baldini}, {Baumgartner}, {Bellazzini}, {Bianchi}, {Bongiorno}, {Bonino}, {Brez}, {Bucciantini}, {Capitanio}, {Castellano}, {Chen}, {Ciprini}, {Costa}, {De Rosa}, {Del Monte}, {Di Lalla}, {Di Marco}, {Donnarumma}, {Doroshenko}, {Dov{\v{c}}iak}, {Ehlert}, {Enoto}, {Evangelista}, {Fabiani}, {Ferrazzoli}, {Garcia}, {Gunji}, {Hayashida}, {Heyl}, {Iwakiri}, {Kaaret}, {Karas},
  {Kislat}, {Kitaguchi}, {Kolodziejczak}, {La Monaca}, {Latronico}, {Maldera}, {Manfreda}, {Marin}, {Marinucci}, {Marshall}, {Massaro}, {Matt}, {Mitsuishi}, {Mizuno}, {Muleri}, {Ng}, {O'Dell}, {Omodei}, {Oppedisano}, {Papitto}, {Pavlov}, {Peirson}, {Perri}, {Pesce-Rollins}, {Petrucci}, {Pilia}, {Possenti}, {Poutanen}, {Puccetti}, {Ramsey}, {Rankin}, {Ratheesh}, {Roberts}, {Sgr{\`o}}, {Slane}, {Soffitta}, {Spandre}, {Swartz}, {Tamagawa}, {Tavecchio}, {Taverna}, {Tawara}, {Tennant}, {Thomas}, {Tombesi}, {Trois}, {Tsygankov}, {Turolla}, {Romani}, {Vink}, {Weisskopf}, {Wu}, {Xie}, \& {Zane}}]{Kouch2024-II}
{Kouch}, P.~M., {Liodakis}, I., {Fenu}, F., {et~al.} 2024{\natexlab{a}}, arXiv e-prints, arXiv:2411.16868.
\newblock \doarXiv{2411.16868}

\bibitem[{{Kouch} {et~al.}(2024{\natexlab{b}}){Kouch}, {Liodakis}, {Middei}, {Kim}, {Tavecchio}, {Marscher}, {Marshall}, {Ehlert}, {Di Gesu}, {Jorstad}, {Agudo}, {Madejski}, {Romani}, {Errando}, {Lindfors}, {Nilsson}, {Toppari}, {Potter}, {Imazawa}, {Sasada}, {Fukazawa}, {Kawabata}, {Uemura}, {Mizuno}, {Nakaoka}, {Akitaya}, {McCall}, {Jermak}, {Steele}, {Myserlis}, {Gurwell}, {Keating}, {Rao}, {Kang}, {Lee}, {Kim}, {Cheong}, {Jeong}, {Angelakis}, {Kraus}, {Jos{\'e} Aceituno}, {Bonnoli}, {Casanova}, {Escudero}, {Ag{\'\i}s-Gonz{\'a}lez}, {Husillos}, {Morcuende}, {Otero-Santos}, {Sota}, {Bachev}, {Antonelli}, {Bachetti}, {Baldini}, {Baumgartner}, {Bellazzini}, {Bianchi}, {Bongiorno}, {Bonino}, {Brez}, {Bucciantini}, {Capitanio}, {Castellano}, {Cavazzuti}, {Chen}, {Ciprini}, {Costa}, {De Rosa}, {Del Monte}, {Di Lalla}, {Di Marco}, {Donnarumma}, {Doroshenko}, {Dov{\v{c}}iak}, {Enoto}, {Evangelista}, {Fabiani}, {Ferrazzoli}, {Garcia}, {Gunji}, {Hayashida}, {Heyl}, {Iwakiri}, {Kaaret}, {Karas}, {Kislat},
  {Kitaguchi}, {Kolodziejczak}, {Krawczynski}, {La Monaca}, {Latronico}, {Maldera}, {Manfreda}, {Marin}, {Marinucci}, {Massaro}, {Matt}, {Mitsuishi}, {Muleri}, {Negro}, {Ng}, {O'Dell}, {Omodei}, {Oppedisano}, {Papitto}, {Pavlov}, {Peirson}, {Perri}, {Pesce-Rollins}, {Petrucci}, {Pilia}, {Possenti}, {Poutanen}, {Puccetti}, {Ramsey}, {Rankin}, {Ratheesh}, {Roberts}, {Sgr{\`o}}, {Slane}, {Soffitta}, {Spandre}, {Swartz}, {Tamagawa}, {Taverna}, {Tawara}, {Tennant}, {Thomas}, {Tombesi}, {Trois}, {Tsygankov}, {Turolla}, {Vink}, {Weisskopf}, {Wu}, {Xie}, \& {Zane}}]{Kouch2024}
{Kouch}, P.~M., {Liodakis}, I., {Middei}, R., {et~al.} 2024{\natexlab{b}}, arXiv e-prints, arXiv:2406.01693, \dodoi{10.48550/arXiv.2406.01693}

\bibitem[{{Kraus} {et~al.}(2003){Kraus}, {Krichbaum}, {Wegner}, {Witzel}, {Cim{\`o}}, {Quirrenbach}, {Britzen}, {Fuhrmann}, {Lobanov}, {Naundorf}, {Otterbein}, {Peng}, {Risse}, {Ros}, \& {Zensus}}]{Krauss2003}
{Kraus}, A., {Krichbaum}, T.~P., {Wegner}, R., {et~al.} 2003, \aap, 401, 161, \dodoi{10.1051/0004-6361:20030118}

\bibitem[{{Krawczynski}(2012)}]{krawczynski2012}
{Krawczynski}, H. 2012, \apj, 744, 30, \dodoi{10.1088/0004-637X/744/1/30}

\bibitem[{{Liodakis} {et~al.}(2019){Liodakis}, {Peirson}, \& {Romani}}]{Liodakis2019}
{Liodakis}, I., {Peirson}, A.~L., \& {Romani}, R.~W. 2019, \apj, 880, 29, \dodoi{10.3847/1538-4357/ab2719}

\bibitem[{{Liodakis} {et~al.}(2022){Liodakis}, {Marscher}, {Agudo}, {Berdyugin}, {Bernardos}, {Bonnoli}, {Borman}, {Casadio}, {Casanova}, {Cavazzuti}, {Rodriguez Cavero}, {Di Gesu}, {Di Lalla}, {Donnarumma}, {Ehlert}, {Errando}, {Escudero}, {Garc{\'\i}a-Comas}, {Ag{\'\i}s-Gonz{\'a}lez}, {Husillos}, {Jormanainen}, {Jorstad}, {Kagitani}, {Kopatskaya}, {Kravtsov}, {Krawczynski}, {Lindfors}, {Larionova}, {Madejski}, {Marin}, {Marchini}, {Marshall}, {Morozova}, {Massaro}, {Masiero}, {Mawet}, {Middei}, {Millar-Blanchaer}, {Myserlis}, {Negro}, {Nilsson}, {O'Dell}, {Omodei}, {Pacciani}, {Paggi}, {Panopoulou}, {Peirson}, {Perri}, {Petrucci}, {Poutanen}, {Puccetti}, {Romani}, {Sakanoi}, {Savchenko}, {Sota}, {Tavecchio}, {Tinyanont}, {Vasilyev}, {Weaver}, {Zhovtan}, {Antonelli}, {Bachetti}, {Baldini}, {Baumgartner}, {Bellazzini}, {Bianchi}, {Bongiorno}, {Bonino}, {Brez}, {Bucciantini}, {Capitanio}, {Castellano}, {Ciprini}, {Costa}, {De Rosa}, {Del Monte}, {Di Marco}, {Doroshenko}, {Dov{\v{c}}iak}, {Enoto},
  {Evangelista}, {Fabiani}, {Ferrazzoli}, {Garcia}, {Gunji}, {Hayashida}, {Heyl}, {Iwakiri}, {Karas}, {Kitaguchi}, {Kolodziejczak}, {La Monaca}, {Latronico}, {Maldera}, {Manfreda}, {Marinucci}, {Matt}, {Mitsuishi}, {Mizuno}, {Muleri}, {Ng}, {Oppedisano}, {Papitto}, {Pavlov}, {Pesce-Rollins}, {Pilia}, {Possenti}, {Ramsey}, {Rankin}, {Ratheesh}, {Sgr{\'o}}, {Slane}, {Soffitta}, {Spandre}, {Tamagawa}, {Taverna}, {Tawara}, {Tennant}, {Thomas}, {Tombesi}, {Trois}, {Tsygankov}, {Turolla}, {Vink}, {Weisskopf}, {Wu}, {Xie}, \& {Zane}}]{liodakis2022}
{Liodakis}, I., {Marscher}, A.~P., {Agudo}, I., {et~al.} 2022, \nat, 611, 677, \dodoi{10.1038/s41586-022-05338-0}

\bibitem[{{Lott} {et~al.}(2020){Lott}, {Gasparrini}, \& {Ciprini}}]{lott2020}
{Lott}, B., {Gasparrini}, D., \& {Ciprini}, S. 2020, arXiv e-prints, arXiv:2010.08406, \dodoi{10.48550/arXiv.2010.08406}

\bibitem[{{Madejski} {et~al.}(1999){Madejski}, {Sikora}, {Jaffe}, {B{\L}a{\.z}ejowski}, {Jahoda}, \& {Moderski}}]{Madejski1999}
{Madejski}, G.~M., {Sikora}, M., {Jaffe}, T., {et~al.} 1999, \apj, 521, 145, \dodoi{10.1086/307524}

\bibitem[{{Maraschi} {et~al.}(1992){Maraschi}, {Ghisellini}, \& {Celotti}}]{maraschi1992}
{Maraschi}, L., {Ghisellini}, G., \& {Celotti}, A. 1992, \apjl, 397, L5, \dodoi{10.1086/186531}

\bibitem[{{Marrone} \& {Rao}(2008)}]{Marrone2008}
{Marrone}, D.~P., \& {Rao}, R. 2008, in \procspie, Vol. 7020, Millimeter and Submillimeter Detectors and Instrumentation for Astronomy IV, ed. W.~D. {Duncan}, W.~S. {Holland}, S.~{Withington}, \& J.~{Zmuidzinas}, 70202B, \dodoi{10.1117/12.788677}

\bibitem[{{Marscher} \& {Gear}(1985)}]{marscher1985}
{Marscher}, A.~P., \& {Gear}, W.~K. 1985, \apj, 298, 114, \dodoi{10.1086/163592}

\bibitem[{{Marscher} \& {Jorstad}(2022)}]{Marscher2022}
{Marscher}, A.~P., \& {Jorstad}, S.~G. 2022, Universe, 8, 644, \dodoi{10.3390/universe8120644}

\bibitem[{{Marscher} {et~al.}(2008){Marscher}, {Jorstad}, {D'Arcangelo}, {Smith}, {Williams}, {Larionov}, {Oh}, {Olmstead}, {Aller}, {Aller}, {McHardy}, {L{\"a}hteenm{\"a}ki}, {Tornikoski}, {Valtaoja}, {Hagen-Thorn}, {Kopatskaya}, {Gear}, {Tosti}, {Kurtanidze}, {Nikolashvili}, {Sigua}, {Miller}, \& {Ryle}}]{Marscher2008}
{Marscher}, A.~P., {Jorstad}, S.~G., {D'Arcangelo}, F.~D., {et~al.} 2008, \nat, 452, 966, \dodoi{10.1038/nature06895}

\bibitem[{{Marshall}(2024)}]{2024ApJ...964...88M}
{Marshall}, H.~L. 2024, \apj, 964, 88, \dodoi{10.3847/1538-4357/ad0897}

\bibitem[{{Marshall} {et~al.}(2023){Marshall}, {Liodakis}, {Marscher}, {Di Lalla}, {Jorstad}, {Kim}, {Middei}, {Negro}, {Omodei}, {Peirson}, {Perri}, {Puccetti}, {Agudo}, {Bonnoli}, {Berdyugin}, {Cavazzuti}, {Rodriguez Cavero}, {Donnarumma}, {Di Gesu}, {Jormanainen}, {Krawczynski}, {Lindfors}, {Marin}, {Massaro}, {Pacciani}, {Poutanen}, {Tavecchio}, {Kouch}, {Aceituno}, {Bernardos}, {Bonnoli}, {Casanova}, {Garcia-Comas}, {Agis-Gonzalez}, {Husillos}, {Marchini}, {Sota}, {Blinov}, {Bourbah}, {Kielhmann}, {Kontopodis}, {Mandarakas}, {Romanopoulos}, {Skalidis}, {Vervelaki}, {Borman}, {Kopatskaya}, {Larionova}, {Morozova}, {Savchenko}, {Vasilyev}, {Zhovtan}, {Casadio}, {Escudero}, {Kramer}, {Myserlis}, {Trainou}, {Imazawa}, {Sasada}, {Fukazawa}, {Kawabata}, {Uemura}, {Mizuno}, {Nakaoka}, {Akitaya}, {Masiero}, {Mawet}, {Millar-Blanchaer}, {Panopoulou}, {Tinyanont}, {Berdyugin}, {Kagitani}, {Kravtsov}, {Sakanoi}, {Antonelli}, {Bachetti}, {Baldini}, {Baumgartner}, {Bellazzini}, {Bianchi}, {Bongiorno}, {Bonino},
  {Brez}, {Bucciantini}, {Capitanio}, {Castellano}, {Cavazzuti}, {Chen}, {Ciprini}, {Costa}, {De Rosa}, {Del Monte}, {Di Gesu}, {Di Marco}, {Donnarumma}, {Doroshenko}, {Dovvciak}, {Ehlert}, {Enoto}, {Evangelista}, {Fabiani}, {Ferrazzoli}, {Garcia}, {Gunji}, {Hayashida}, {Heyl}, {Iwakiri}, {Kaaret}, {Karas}, {Kislat}, {Kitaguchi}, {Kolodziejczak}, {Krawczynski}, {La Monaca}, {Latronico}, {Maldera}, {Manfreda}, {Marin}, {Marinucci}, {Matt}, {Mitsuishi}, {Mizuno}, {Muleri}, {Ng}, {ODell}, {Oppedisano}, {Papitto}, {Pavlov}, {Pesce-Rollins}, {Petrucci}, {Pilia}, {Possenti}, {Poutanen}, {Puccetti}, {Ramsey}, {Rankin}, {Ratheesh}, {Roberts}, {Romani}, {Sgro}, {Slane}, {Soffitta}, {Spandre}, {Swartz}, {Tamagawa}, {Taverna}, {Tawara}, {Tennant}, {Thomas}, {Tombesi}, {Trois}, {Tsygankov}, {Turolla}, {Vink}, {Weisskopf}, {Wu}, {Xie}, \& {Zane}}]{Marshall2023}
{Marshall}, H.~L., {Liodakis}, I., {Marscher}, A.~P., {et~al.} 2023, arXiv e-prints, arXiv:2310.11510, \dodoi{10.48550/arXiv.2310.11510}

\bibitem[{{Middei} {et~al.}(2023){Middei}, {Liodakis}, {Perri}, {Puccetti}, {Cavazzuti}, {Di Gesu}, {Ehlert}, {Madejski}, {Marscher}, {Marshall}, {Muleri}, {Negro}, {Jorstad}, {Ag{\'\i}s-Gonz{\'a}lez}, {Agudo}, {Bonnoli}, {Bernardos}, {Casanova}, {Garc{\'\i}a-Comas}, {Husillos}, {Marchini}, {Sota}, {Kouch}, {Lindfors}, {Borman}, {Kopatskaya}, {Larionova}, {Morozova}, {Savchenko}, {Vasilyev}, {Zhovtan}, {Casadio}, {Escudero}, {Myserlis}, {Hales}, {Kameno}, {Kneissl}, {Messias}, {Nagai}, {Blinov}, {Bourbah}, {Kiehlmann}, {Kontopodis}, {Mandarakas}, {Romanopoulos}, {Skalidis}, {Vervelaki}, {Masiero}, {Mawet}, {Millar-Blanchaer}, {Panopoulou}, {Tinyanont}, {Berdyugin}, {Kagitani}, {Kravtsov}, {Sakanoi}, {Imazawa}, {Sasada}, {Fukazawa}, {Kawabata}, {Uemura}, {Mizuno}, {Nakaoka}, {Akitaya}, {Gurwell}, {Rao}, {Di Lalla}, {Cibrario}, {Donnarumma}, {Kim}, {Omodei}, {Pacciani}, {Poutanen}, {Tavecchio}, {Antonelli}, {Bachetti}, {Baldini}, {Baumgartner}, {Bellazzini}, {Bianchi}, {Bongiorno}, {Bonino}, {Brez},
  {Bucciantini}, {Capitanio}, {Castellano}, {Ciprini}, {Costa}, {De Rosa}, {Del Monte}, {Di Marco}, {Doroshenko}, {Dov{\v{c}}iak}, {Enoto}, {Evangelista}, {Fabiani}, {Ferrazzoli}, {Garcia}, {Gunji}, {Hayashida}, {Heyl}, {Iwakiri}, {Karas}, {Kitaguchi}, {Kolodziejczak}, {Krawczynski}, {La Monaca}, {Latronico}, {Maldera}, {Manfreda}, {Marin}, {Marinucci}, {Massaro}, {Matt}, {Mitsuishi}, {Ng}, {O'Dell}, {Oppedisano}, {Papitto}, {Pavlov}, {Peirson}, {Pesce-Rollins}, {Petrucci}, {Pilia}, {Possenti}, {Ramsey}, {Rankin}, {Ratheesh}, {Romani}, {Sgr{\'o}}, {Slane}, {Soffitta}, {Spandre}, {Tamagawa}, {Taverna}, {Tawara}, {Tennant}, {Thomas}, {Tombesi}, {Trois}, {Tsygankov}, {Turolla}, {Vink}, {Weisskopf}, {Wu}, {Xie}, \& {Zane}}]{middei2023}
{Middei}, R., {Liodakis}, I., {Perri}, M., {et~al.} 2023, \apjl, 942, L10, \dodoi{10.3847/2041-8213/aca281}

\bibitem[{{Mondal} {et~al.}(2025){Mondal}, {Sar}, {Kundu}, {Chatterjee}, \& {Majumdar}}]{2025ApJ...978...43M}
{Mondal}, A., {Sar}, A., {Kundu}, M., {Chatterjee}, R., \& {Majumdar}, P. 2025, \apj, 978, 43, \dodoi{10.3847/1538-4357/ad9603}

\bibitem[{{Myserlis} {et~al.}(2018){Myserlis}, {Angelakis}, {Kraus}, {Liontas}, {Marchili}, {Aller}, {Aller}, {Karamanavis}, {Fuhrmann}, {Krichbaum}, \& {Zensus}}]{Myserlis2018}
{Myserlis}, I., {Angelakis}, E., {Kraus}, A., {et~al.} 2018, \aap, 609, A68, \dodoi{10.1051/0004-6361/201630301}

\bibitem[{{Nagirner} \& {Poutanen}(1993)}]{Nagirner1993}
{Nagirner}, D.~I., \& {Poutanen}, J. 1993, \aap, 275, 325

\bibitem[{{Nasa High Energy Astrophysics Science Archive Research Center (Heasarc)}(2014)}]{2014ascl.soft08004N}
{Nasa High Energy Astrophysics Science Archive Research Center (Heasarc)}. 2014, {HEAsoft: Unified Release of FTOOLS and XANADU}, Astrophysics Source Code Library, record ascl:1408.004

\bibitem[{{Nilsson} {et~al.}(2007){Nilsson}, {Pasanen}, {Takalo}, {Lindfors}, {Berdyugin}, {Ciprini}, \& {Pforr}}]{Nilsson2007}
{Nilsson}, K., {Pasanen}, M., {Takalo}, L.~O., {et~al.} 2007, \aap, 475, 199, \dodoi{10.1051/0004-6361:20077624}

\bibitem[{{Nilsson} {et~al.}(2018){Nilsson}, {Lindfors}, {Takalo}, {Reinthal}, {Berdyugin}, {Sillanp{\"a}{\"a}}, {Ciprini}, {Halkola}, {Hein{\"a}m{\"a}ki}, {Hovatta}, {Kadenius}, {Nurmi}, {Ostorero}, {Pasanen}, {Rekola}, {Saarinen}, {Sainio}, {Tuominen}, {Villforth}, {Vornanen}, \& {Zaprudin}}]{Nilsson2018}
{Nilsson}, K., {Lindfors}, E., {Takalo}, L.~O., {et~al.} 2018, \aap, 620, A185, \dodoi{10.1051/0004-6361/201833621}

\bibitem[{{Otero-Santos} {et~al.}(2024){Otero-Santos}, {Piirola}, {Escudero Pedrosa}, {Agudo}, {Morcuende}, {Sota}, {Casanova}, {Aceituno}, \& {Santos-Sanz}}]{otero2024}
{Otero-Santos}, J., {Piirola}, V., {Escudero Pedrosa}, J., {et~al.} 2024, \aj, 167, 137, \dodoi{10.3847/1538-3881/ad250d}

\bibitem[{{Paliya} {et~al.}(2018){Paliya}, {Zhang}, {B{\"o}ttcher}, {Ajello}, {Dom{\'\i}nguez}, {Joshi}, {Hartmann}, \& {Stalin}}]{Paliya2018}
{Paliya}, V.~S., {Zhang}, H., {B{\"o}ttcher}, M., {et~al.} 2018, \apj, 863, 98, \dodoi{10.3847/1538-4357/aad1f0}

\bibitem[{{Panopoulou} {et~al.}(2015){Panopoulou}, {Tassis}, {Blinov}, {Pavlidou}, {King}, {Paleologou}, {Ramaprakash}, {Angelakis}, {Balokovi{\'c}}, {Das}, {Feiler}, {Hovatta}, {Khodade}, {Kiehlmann}, {Kus}, {Kylafis}, {Liodakis}, {Mahabal}, {Modi}, {Myserlis}, {Papadakis}, {Papamastorakis}, {Pazderska}, {Pazderski}, {Pearson}, {Rajarshi}, {Readhead}, {Reig}, \& {Zensus}}]{Panopoulou2015}
{Panopoulou}, G., {Tassis}, K., {Blinov}, D., {et~al.} 2015, \mnras, 452, 715, \dodoi{10.1093/mnras/stv1301}

\bibitem[{{Peirson} {et~al.}(2022){Peirson}, {Liodakis}, \& {Romani}}]{Peirson2022}
{Peirson}, A.~L., {Liodakis}, I., \& {Romani}, R.~W. 2022, \apj, 931, 59, \dodoi{10.3847/1538-4357/ac6a54}

\bibitem[{{Peirson} \& {Romani}(2019)}]{peirson2019}
{Peirson}, A.~L., \& {Romani}, R.~W. 2019, \apj, 885, 76, \dodoi{10.3847/1538-4357/ab46b1}

\bibitem[{{Peirson} {et~al.}(2023){Peirson}, {Negro}, {Liodakis}, {Middei}, {Kim}, {Marscher}, {Marshall}, {Pacciani}, {Romani}, {Wu}, {Di Marco}, {Di Lalla}, {Omodei}, {Jorstad}, {Agudo}, {Kouch}, {Lindfors}, {Aceituno}, {Bernardos}, {Bonnoli}, {Casanova}, {Garc{\'\i}a-Comas}, {Ag{\'\i}s-Gonz{\'a}lez}, {Husillos}, {Marchini}, {Sota}, {Casadio}, {Escudero}, {Myserlis}, {Sievers}, {Gurwell}, {Rao}, {Imazawa}, {Sasada}, {Fukazawa}, {Kawabata}, {Uemura}, {Mizuno}, {Nakaoka}, {Akitaya}, {Cheong}, {Jeong}, {Kang}, {Kim}, {Lee}, {Angelakis}, {Kraus}, {Cibrario}, {Donnarumma}, {Poutanen}, {Tavecchio}, {Antonelli}, {Bachetti}, {Baldini}, {Baumgartner}, {Bellazzini}, {Bianchi}, {Bongiorno}, {Bonino}, {Brez}, {Bucciantini}, {Capitanio}, {Castellano}, {Cavazzuti}, {Chen}, {Ciprini}, {Costa}, {De Rosa}, {Del Monte}, {Di Gesu}, {Doroshenko}, {Dov{\v{c}}iak}, {Ehlert}, {Enoto}, {Evangelista}, {Fabiani}, {Ferrazzoli}, {Garcia}, {Gunji}, {Hayashida}, {Heyl}, {Iwakiri}, {Kaaret}, {Karas}, {Kitaguchi}, {Kolodziejczak},
  {Krawczynski}, {La Monaca}, {Latronico}, {Madejski}, {Maldera}, {Manfreda}, {Marin}, {Marinucci}, {Massaro}, {Matt}, {Mitsuishi}, {Muleri}, {Ng}, {O'Dell}, {Oppedisano}, {Papitto}, {Pavlov}, {Perri}, {Pesce-Rollins}, {Petrucci}, {Pilia}, {Possenti}, {Puccetti}, {Ramsey}, {Rankin}, {Ratheesh}, {Roberts}, {Sgr{\'o}}, {Slane}, {Soffitta}, {Spandre}, {Swartz}, {Tamagawa}, {Taverna}, {Tawara}, {Tennant}, {Thomas}, {Tombesi}, {Trois}, {Tsygankov}, {Turolla}, {Vink}, {Weisskopf}, {Xie}, \& {Zane}}]{peirson2023}
{Peirson}, A.~L., {Negro}, M., {Liodakis}, I., {et~al.} 2023, \apjl, 948, L25, \dodoi{10.3847/2041-8213/acd242}

\bibitem[{{Pesce-Rollins} {et~al.}(2019){Pesce-Rollins}, {Lalla}, {Omodei}, \& {Baldini}}]{pesce2019}
{Pesce-Rollins}, M., {Lalla}, N.~D., {Omodei}, N., \& {Baldini}, L. 2019, Nuclear Instruments and Methods in Physics Research A, 936, 224, \dodoi{10.1016/j.nima.2018.10.041}

\bibitem[{{Piconcelli} {et~al.}(2004){Piconcelli}, {Jimenez-Bail{\'o}n}, {Guainazzi}, {Schartel}, {Rodr{\'\i}guez-Pascual}, \& {Santos-Lle{\'o}}}]{Piconcelli2004}
{Piconcelli}, E., {Jimenez-Bail{\'o}n}, E., {Guainazzi}, M., {et~al.} 2004, \mnras, 351, 161, \dodoi{10.1111/j.1365-2966.2004.07764.x}

\bibitem[{{Poutanen}(1994)}]{Poutanen1994}
{Poutanen}, J. 1994, \apjs, 92, 607, \dodoi{10.1086/192024}

\bibitem[{{Primiani} {et~al.}(2016){Primiani}, {Young}, {Young}, {Patel}, {Wilson}, {Vertatschitsch}, {Chitwood}, {Srinivasan}, {MacMahon}, \& {Weintroub}}]{Primiani2016}
{Primiani}, R.~A., {Young}, K.~H., {Young}, A., {et~al.} 2016, Journal of Astronomical Instrumentation, 5, 1641006, \dodoi{10.1142/S2251171716410063}

\bibitem[{{Ramaprakash} {et~al.}(2019){Ramaprakash}, {Rajarshi}, {Das}, {Khodade}, {Modi}, {Panopoulou}, {Maharana}, {Blinov}, {Angelakis}, {Casadio}, {Fuhrmann}, {Hovatta}, {Kiehlmann}, {King}, {Kylafis}, {Kougentakis}, {Kus}, {Mahabal}, {Marecki}, {Myserlis}, {Paterakis}, {Paleologou}, {Liodakis}, {Papadakis}, {Papamastorakis}, {Pavlidou}, {Pazderski}, {Pearson}, {Readhead}, {Reig}, {S{\l}owikowska}, {Tassis}, \& {Zensus}}]{Ramaprakash2019}
{Ramaprakash}, A.~N., {Rajarshi}, C.~V., {Das}, H.~K., {et~al.} 2019, \mnras, 485, 2355, \dodoi{10.1093/mnras/stz557}

\bibitem[{{Ripperda} {et~al.}(2022){Ripperda}, {Liska}, {Chatterjee}, {Musoke}, {Philippov}, {Markoff}, {Tchekhovskoy}, \& {Younsi}}]{Ripperda2022}
{Ripperda}, B., {Liska}, M., {Chatterjee}, K., {et~al.} 2022, \apjl, 924, L32, \dodoi{10.3847/2041-8213/ac46a1}

\bibitem[{{Roming} {et~al.}(2005){Roming}, {Kennedy}, {Mason}, {Nousek}, {Ahr}, {Bingham}, {Broos}, {Carter}, {Hancock}, {Huckle}, {Hunsberger}, {Kawakami}, {Killough}, {Koch}, {McLelland}, {Smith}, {Smith}, {Soto}, {Boyd}, {Breeveld}, {Holland}, {Ivanushkina}, {Pryzby}, {Still}, \& {Stock}}]{Rom05}
{Roming}, P. W.~A., {Kennedy}, T.~E., {Mason}, K.~O., {et~al.} 2005, \ssr, 120, 95, \dodoi{10.1007/s11214-005-5095-4}

\bibitem[{{Shepherd}(1997)}]{1997ASPC..125...77S}
{Shepherd}, M.~C. 1997, in Astronomical Society of the Pacific Conference Series, Vol. 125, Astronomical Data Analysis Software and Systems VI, ed. G.~{Hunt} \& H.~{Payne}, 77

\bibitem[{{Shrestha} {et~al.}(2020){Shrestha}, {Steele}, {Piascik}, {Jermak}, {Smith}, \& {Copperwheat}}]{Shrestha2020}
{Shrestha}, M., {Steele}, I.~A., {Piascik}, A.~S., {et~al.} 2020, \mnras, 494, 4676, \dodoi{10.1093/mnras/staa1049}

\bibitem[{{Smith}(2017)}]{Smith2017}
{Smith}, P.~S. 2017, The Astronomer's Telegram, 11047, 1

\bibitem[{{Smith} {et~al.}(2009){Smith}, {Montiel}, {Rightley}, {Turner}, {Schmidt}, \& {Jannuzi}}]{smith2009}
{Smith}, P.~S., {Montiel}, E., {Rightley}, S., {et~al.} 2009, 2009 Fermi Symposium, eConf Proceedings C091122, arXiv:0912.3621, \dodoi{10.48550/arXiv.0912.3621}

\bibitem[{{Str{\"u}der} {et~al.}(2001){Str{\"u}der}, {Briel}, {Dennerl}, {Hartmann}, {Kendziorra}, {Meidinger}, {Pfeffermann}, {Reppin}, {Aschenbach}, {Bornemann}, {Br{\"a}uninger}, {Burkert}, {Elender}, {Freyberg}, {Haberl}, {Hartner}, {Heuschmann}, {Hippmann}, {Kastelic}, {Kemmer}, {Kettenring}, {Kink}, {Krause}, {M{\"u}ller}, {Oppitz}, {Pietsch}, {Popp}, {Predehl}, {Read}, {Stephan}, {St{\"o}tter}, {Tr{\"u}mper}, {Holl}, {Kemmer}, {Soltau}, {St{\"o}tter}, {Weber}, {Weichert}, {von Zanthier}, {Carathanassis}, {Lutz}, {Richter}, {Solc}, {B{\"o}ttcher}, {Kuster}, {Staubert}, {Abbey}, {Holland}, {Turner}, {Balasini}, {Bignami}, {La Palombara}, {Villa}, {Buttler}, {Gianini}, {Lain{\'e}}, {Lumb}, \& {Dhez}}]{Struder2001}
{Str{\"u}der}, L., {Briel}, U., {Dennerl}, K., {et~al.} 2001, \aap, 365, L18, \dodoi{10.1051/0004-6361:20000066}

\bibitem[{{Telescope Array Collaboration} {et~al.}(2023){Telescope Array Collaboration}, {Abbasi}, {Allen}, {Arimura}, {Belz}, {Bergman}, {Blake}, {Shin}, {Buckland}, {Cheon}, {Fujii}, {Fujisue}, {Fujita}, {Fukushima}, {Furlich}, {Gerber}, {Globus}, {Hibino}, {Higuchi}, {Honda}, {Ikeda}, {Ito}, {Iwasaki}, {Jeong}, {Jeong}, {Jui}, {Kadota}, {Kakimoto}, {Kalashev}, {Kasahara}, {Kawata}, {Kharuk}, {Kido}, {Kim}, {Kim}, {Kim}, {Kim}, {Komae}, {Kubota}, {Kuznetsov}, {Lee}, {Lubsandorzhiev}, {Lundquist}, {Matthews}, {Nagataki}, {Nakamura}, {Nakazawa}, {Nonaka}, {Ogio}, {Ono}, {Oshima}, {Park}, {Potts}, {Pshirkov}, {Remington}, {Rodriguez}, {Rott}, {Rubtsov}, {Ryu}, {Sagawa}, {Sakaki}, {Sako}, {Sakurai}, {Shin}, {Smith}, {Sokolsky}, {Stokes}, {Stroman}, {Takahashi}, {Takeda}, {Taketa}, {Tameda}, {Thomas}, {Thomson}, {Tinyakov}, {Tkachev}, {Tomida}, {Troitsky}, {Tsunesada}, {Udo}, {Urban}, {Wong}, {Yamazaki}, {Yuma}, {Zhezher}, \& {Zundel}}]{Amaterasu2023}
{Telescope Array Collaboration}, {Abbasi}, R.~U., {Allen}, M.~G., {et~al.} 2023, Science, 382, 903, \dodoi{10.1126/science.abo5095}

\bibitem[{{Tramacere}(2020)}]{tramacere2020}
{Tramacere}, A. 2020, {JetSeT: Numerical modeling and SED fitting tool for relativistic jets}, Astrophysics Source Code Library, record ascl:2009.001

\bibitem[{{Tramacere} {et~al.}(2009){Tramacere}, {Giommi}, {Perri}, {Verrecchia}, \& {Tosti}}]{tramacere2009}
{Tramacere}, A., {Giommi}, P., {Perri}, M., {Verrecchia}, F., \& {Tosti}, G. 2009, \aap, 501, 879, \dodoi{10.1051/0004-6361/200810865}

\bibitem[{{Tramacere} {et~al.}(2011){Tramacere}, {Massaro}, \& {Taylor}}]{tramacere2011}
{Tramacere}, A., {Massaro}, E., \& {Taylor}, A.~M. 2011, \apj, 739, 66, \dodoi{10.1088/0004-637X/739/2/66}

\bibitem[{{Weaver} {et~al.}(2020){Weaver}, {Williamson}, {Jorstad}, {Marscher}, {Larionov}, {Raiteri}, {Villata}, {Acosta-Pulido}, {Bachev}, {Baida}, {Balonek}, {Ben{\'\i}tez}, {Borman}, {Bozhilov}, {Carnerero}, {Carosati}, {Chen}, {Damljanovic}, {Dhiman}, {Dougherty}, {Ehgamberdiev}, {Grishina}, {Gupta}, {Hart}, {Hiriart}, {Hsiao}, {Ibryamov}, {Joner}, {Kimeridze}, {Kopatskaya}, {Kurtanidze}, {Kurtanidze}, {Larionova}, {Matsumoto}, {Matsumura}, {Minev}, {Mirzaqulov}, {Morozova}, {Nikiforova}, {Nikolashvili}, {Ovcharov}, {Rizzi}, {Sadun}, {Savchenko}, {Semkov}, {Slater}, {Smith}, {Stojanovic}, {Strigachev}, {Troitskaya}, {Troitsky}, {Tsai}, {Vince}, {Valcheva}, {Vasilyev}, {Zaharieva}, \& {Zhovtan}}]{Weaver2020}
{Weaver}, Z.~R., {Williamson}, K.~E., {Jorstad}, S.~G., {et~al.} 2020, \apj, 900, 137, \dodoi{10.3847/1538-4357/aba693}

\bibitem[{{Weaver} {et~al.}(2022){Weaver}, {Jorstad}, {Marscher}, {Morozova}, {Troitsky}, {Agudo}, {G{\'o}mez}, {L{\"a}hteenm{\"a}ki}, {Tammi}, \& {Tornikoski}}]{Weaver2022}
{Weaver}, Z.~R., {Jorstad}, S.~G., {Marscher}, A.~P., {et~al.} 2022, \apjs, 260, 12, \dodoi{10.3847/1538-4365/ac589c}

\bibitem[{{Weisskopf} {et~al.}(2022){Weisskopf}, {Soffitta}, {Baldini}, {Ramsey}, {O'Dell}, {Romani}, {Matt}, {Deininger}, {Baumgartner}, {Bellazzini}, {Costa}, {Kolodziejczak}, {Latronico}, {Marshall}, {Muleri}, {Bongiorno}, {Tennant}, {Bucciantini}, {Dovciak}, {Marin}, {Marscher}, {Poutanen}, {Slane}, {Turolla}, {Kalinowski}, {Di Marco}, {Fabiani}, {Minuti}, {La Monaca}, {Pinchera}, {Rankin}, {Sgro'}, {Trois}, {Xie}, {Alexander}, {Allen}, {Amici}, {Andersen}, {Antonelli}, {Antoniak}, {Attin{\`a}}, {Barbanera}, {Bachetti}, {Baggett}, {Bladt}, {Brez}, {Bonino}, {Boree}, {Borotto}, {Breeding}, {Brienza}, {Bygott}, {Caporale}, {Cardelli}, {Carpentiero}, {Castellano}, {Castronuovo}, {Cavalli}, {Cavazzuti}, {Ceccanti}, {Centrone}, {Citraro}, {D'Amico}, {D'Alba}, {Di Gesu}, {Del Monte}, {Dietz}, {Di Lalla}, {Persio}, {Dolan}, {Donnarumma}, {Evangelista}, {Ferrant}, {Ferrazzoli}, {Ferrie}, {Footdale}, {Forsyth}, {Foster}, {Garelick}, {Gunji}, {Gurnee}, {Head}, {Hibbard}, {Johnson}, {Kelly}, {Kilaru}, {Lefevre},
  {Roy}, {Loffredo}, {Lorenzi}, {Lucchesi}, {Maddox}, {Magazzu}, {Maldera}, {Manfreda}, {Mangraviti}, {Marengo}, {Marrocchesi}, {Massaro}, {Mauger}, {McCracken}, {McEachen}, {Mize}, {Mereu}, {Mitchell}, {Mitsuishi}, {Morbidini}, {Mosti}, {Nasimi}, {Negri}, {Negro}, {Nguyen}, {Nitschke}, {Nuti}, {Onizuka}, {Oppedisano}, {Orsini}, {Osborne}, {Pacheco}, {Paggi}, {Painter}, {Pavelitz}, {Pentz}, {Piazzolla}, {Perri}, {Pesce-Rollins}, {Peterson}, {Pilia}, {Profeti}, {Puccetti}, {Ranganathan}, {Ratheesh}, {Reedy}, {Root}, {Rubini}, {Ruswick}, {Sanchez}, {Sarra}, {Santoli}, {Scalise}, {Sciortino}, {Schroeder}, {Seek}, {Sosdian}, {Spandre}, {Speegle}, {Tamagawa}, {Tardiola}, {Tobia}, {Thomas}, {Valerie}, {Vimercati}, {Walden}, {Weddendorf}, {Wedmore}, {Welch}, {Zanetti}, \& {Zanetti}}]{Weisskopf2022_ixpe_technical}
{Weisskopf}, M.~C., {Soffitta}, P., {Baldini}, L., {et~al.} 2022, Journal of Astronomical Telescopes, Instruments, and Systems, 8, 026002, \dodoi{10.1117/1.JATIS.8.2.026002}

\bibitem[{{Yang} {et~al.}(2024){Yang}, {Yuan}, {Li}, {Mizuno}, {Guo}, {Lu}, {Ho}, {Lin}, {Zdziarski}, \& {Wang}}]{Yang2024}
{Yang}, H., {Yuan}, F., {Li}, H., {et~al.} 2024, Science Advances, 10, eadn3544, \dodoi{10.1126/sciadv.adn3544}

\bibitem[{{Zhang} \& {B{\"o}ttcher}(2013)}]{zhang2013}
{Zhang}, H., \& {B{\"o}ttcher}, M. 2013, \apj, 774, 18, \dodoi{10.1088/0004-637X/774/1/18}

\bibitem[{{Zhang} {et~al.}(2024){Zhang}, {B{\"o}ttcher}, \& {Liodakis}}]{Zhang2024}
{Zhang}, H., {B{\"o}ttcher}, M., \& {Liodakis}, I. 2024, arXiv e-prints, arXiv:2404.12475, \dodoi{10.48550/arXiv.2404.12475}

\bibitem[{{Zhang} {et~al.}(2016){Zhang}, {Deng}, {Li}, \& {B{\"o}ttcher}}]{Zhang2016}
{Zhang}, H., {Deng}, W., {Li}, H., \& {B{\"o}ttcher}, M. 2016, \apj, 817, 63, \dodoi{10.3847/0004-637X/817/1/63}

\bibitem[{{Zhang} {et~al.}(2019){Zhang}, {Fang}, {Li}, {Giannios}, {B{\"o}ttcher}, \& {Buson}}]{Zhang2019}
{Zhang}, H., {Fang}, K., {Li}, H., {et~al.} 2019, \apj, 876, 109, \dodoi{10.3847/1538-4357/ab158d}

\bibitem[{{Zhang} {et~al.}(2017){Zhang}, {Li}, {Guo}, \& {Taylor}}]{Zhang2017}
{Zhang}, H., {Li}, H., {Guo}, F., \& {Taylor}, G. 2017, \apj, 835, 125, \dodoi{10.3847/1538-4357/835/2/125}

\end{thebibliography}

\bibliographystyle{aasjournal}



\end{document}